\documentclass[apj]{emulateapj}
 \usepackage{multirow}
 \usepackage{natbib}
 \usepackage{graphicx}
 \usepackage{graphics}
 \usepackage{epstopdf}
 \usepackage{amsmath}
\newcommand{\chandra}{{\sl Chandra }}
\newcommand{\suzaku}{{\sl Suzaku }}
\newcommand{\wise}{{\sl WISE}}
\newcommand{\allwise}{{\sl ALLWISE }}
\newcommand{\iras}{{\sl IRAS}}
\newcommand{\tmass}{{\sl 2MASS }}
\newcommand{\newton}{{\sl XMM-Newton }}
\newcommand{\sfr}{{\rm SFR}}
\newcommand{\sfrs}{{\rm SFRs }}

\begin{document}
\nocite{*}
\title{Chandra Survey of Nearby Highly Inclined Disk Galaxies - V: Emission Structure and Origin of Galactic Coronae}
\author{Xiaochuan Jiang\altaffilmark{1}, Jiangtao Li\altaffilmark{2}, Taotao Fang\altaffilmark{1*}, Q. Daniel Wang\altaffilmark{3}}
\altaffiltext{1}{Department of Astronomy and Institute of Theoretical Physics and Astrophysics, Xiamen University; fangt@xmu.edu.cn}
\altaffiltext{2}{Department of Astronomy, University of Michigan, 311 West Hall, 1085 S. University Ave, Ann Arbor, MI 48109-1107, USA}
\altaffiltext{3}{Department of Astronomy, University of Massachusetts, 710 North Pleasant St, Amherst, MA 01003, USA}
%\altaffiltext{4}{School of Astronomy and Space Science, Nanjing University, Nanjiang 210093, China}


\begin{abstract}
	The origin of the extended soft X-ray emission around nearby highly inclined disk galaxies (often called as X-ray corona) remains uncertain. The emission could arise from volume-filling hot gas and/or its interaction with cool gas. Morphological properties of the X-ray emission can provide additional information to distinguish these different origins. We define model-independent parameters $H_{50}$, $H_{75}$, and $H_{95}$ --- vertical scales that enclose 50\%, 75\%, and 95\% of the total flux of the emission, respectively. We study the correlation of these parameters with galaxy properties inferred from infrared observations of a sample of nearby highly inclined disk galaxies with high quality Chandra data.
	We find weak $\it negative$ correlations between $H_{50}$ or $H_{75}$ and the surface star formation rate ($I_{\sfr}$), and no correlation for $H_{95}$. However, we detect strong, $\it negative$ correlations of the vertical concentration of the emission, defined as $H_{50}/H_{95}$ or $H_{75}/H_{95}$, with $I_{\sfr}$. Our findings suggest that the X-ray emission around disk galaxies is likely comprised of two components: the extended, weak emission, characterized by $H_{95}$, is influenced by the outflowing hot gas entrained in star formation driven winds, whereas the strong emission close to the disk which is often rich in cool gas characterized by $H_{50}$ or $H_{75}$, is largely impacted by cool-hot gas interaction. 

\end{abstract}

\section{Introduction} The galactic corona represents the hot, ionized, and rarefied gas with $T\sim10^6 K$ in the halo of the Milky way (MW) or other galaxies. Such a corona, traced by diffuse soft X-ray emission, serves as a reservoir of baryons that are either accreted from the intergalactic medium (IGM), or ejected from galaxies (e.g. \citealt{White1978,White1991,Fukugita2006,Bregman2018}). In the latter case, the metal-rich gas is ejected from galaxies by energetic feedback from stellar sources or active galactic nuclei (AGN; e.g. \citealt{Bower2012,Hopkins2014,Agertz2015,Keller2015,Mitra2015}). The X-ray luminosity and morphology of the corona are largely determined by the feedback. Conversely, X-ray observations of galactic coronae provide a potentially powerful tool to probe various poorly understood processes involved in the feedback.
%These baryons derive from both the metal-poor gas accreted from the intergalactic medium IGM and the metal-rich gas which is ejected 

The diffuse soft X-ray emission of the hot corona has been detected by modern X-ray telescopes, such as \chandra, \newton, and \suzaku around nearby galaxies (e.g., \citealt{Wang2001,Strickland2004,Tuellmann2006,Li2008,Wang2010,Li2013,Li2013a,Yamasaki2009}).
%These galaxies include elliptical galaxies containing AGN, normal disk and dwarf galaxies with low/high star formation, and interacting or merging galaxies.
%The emission becomes a good tracer to explore the link between the hot gas and the galaxy properties (e.g. \citealt{Strickland2002}).
%The X-ray emission presented in various type galaxies indicates that the corona may be a common phenomenon in the disk galaxies and play an important role in galaxy formation.
%
%There are two major scenarios proposed for the origin of the corona, one is the accretion of the IGM, and the other one is the feedback processes, such as AGN and SF feedback.
There are tight correlations between the coronal X-ray luminosity ($L_X$) and the global properties of the galaxies (e.g., $M_*$, \sfr, etc. \citealt{Tuellmann2006,Li2008,Li2013a}). For active star forming (SF) galaxies, the coronal luminosity is linearly correlated with the star formation rate (\sfr; e.g. \citealt{Strickland2004,Grimes2005}) and the metallicity of the coronal gas shows the same patterns as being enriched by young stellar populations such as massive stars or core collapsed SNe (e.g., \citealt{Martin2002,Yamasaki2009,Konami2011}). 
For SF-quiescent elliptical galaxies, in contrast, the X-ray luminosity is tightly correlated with the stellar mass ($M_*$), and the slope of the correlation is far larger than unity, especially for massive galaxies (\citealt{Forman1985,Canizares1987,Helsdon2001,Mathews2003,OSullivan2003,Kim2013,Forbes2016}). This strong correlation suggests that gravity plays an important role in heating and/or confining the hot gas.
%However, in modern galaxy formation theory, the gravity or the mass of the dark matter halo of the galaxy not only closely links to the galaxy stellar mass or the \sfr, but also determines the radiative cooling rate or X-ray emission of the accreted gas (\citealt{Dave2008,Dave2011}). %Therefore, no matter how the corona is heated, the positive correlation is expected to present among the coronal soft X-ray luminosity, the stellar mass, the \sfr, and the halo mass of galaxies.
The above correlations could be largely affected by the global scaling relations between various parameters of the galaxies and the overall size of the galaxy.
In the previous work, \citet{Wang2016} revealed a clearly sublinear correlation between the specific X-ray luminosity $L_X/L_k$, where $L_k$ is the K-band luminosity, and the SF intensity ($I_{\sfr}$). The well known linear correlation between $L_X$ and \sfr~is thus largely due to the global scaling of both parameters with the galaxy size. The anti-correlation between $L_X/\sfr$ and $I_\sfr$ further suggests that the bulk of the mechanical energy from stellar feedback is most likely dissipated far away from the galaxy disc with reduced global X-ray emission rate as the SF intensity increases.

Despite the progress, the exact origin of the extraplanar soft X-ray emission remains unclear. In addition to the optically thin volume filling thermal plasma, the charge exchange X-ray emission (CXE), generated at the interface between the cool and hot gas, may also make non-negligible contribution (e.g., \citealt{Liu2011,Liu2012,Zhang2014,Henley2015}). The CXE tends to dominate only in a few special emission lines which are hardly separated from emission lines primarily produced in thermal plasma in low resolution CCD spectra. %It is charactered by the cool-hot interface area other than the amount of cooling gas. 
In many cases, the cool gas traced by $H_\alpha$ emission spatially coincides with the X-ray emission, indicating significant contributions of the CXE (\citealt{Lallement2004}).
Moreover, a high-temperature ($T\gtrsim10^7K$) thermal component, which has low density and hence low emissivity, has been suggested based on the detection of the highly ionized Fe K lines in the nuclear region of M82 (\citealt{Strickland2009}). 
Such a high-temperature gas, if exists, may produce strong soft X-ray emission via interacting with the pre-existing cool gas in and around the galaxies (\citealt{Strickland2000,Strickland2000a}). Statistical analysis of the extraplanar emission morphology may help to separate these different soft X-ray emission components (thermal plasma versus cool-hot gas interaction), as they are expected to have different spatial distributions.

Infrared is an ideal band to explore the stellar mass distribution and the obscured star formation history, particular for the highly inclined galaxies which are heavily affected by the interstellar dust extinction.
The Wide field Infrared Survey Explorer (\wise) mapped the full sky at mid-infrared four bands 3.4 \micron~(W1), 4.6 \micron~(W2), 12 \micron~(W3), and 22 \micron~(W4) with angular resolutions of 6.1\arcsec, 6.4\arcsec, 6.5\arcsec, and 12.0\arcsec, respectively (\citealt{Wright2010}). The sensitivity of the \wise~is much higher than the previous mission, the Infrared Astronomical Satellite (\iras). W1 and W2 of \wise, trace the stellar mass constant, and the other two bands, W3 and W4, are sensitive to the star formation.
In this paper, we compare the mid-infrared galaxy properties obtained from \wise~to the soft X-ray morphology of the extraplanar emission characterized with different parameters. 
We describe our sample and present the data analysis in Section 2. The result and discussion are presented in Section 3 and 4, respectively. Finally, we give our summary in Section 5.

\section{Sample and Data}

 \begin{figure*}[h]
   \center
    \includegraphics[width=0.8\textwidth]{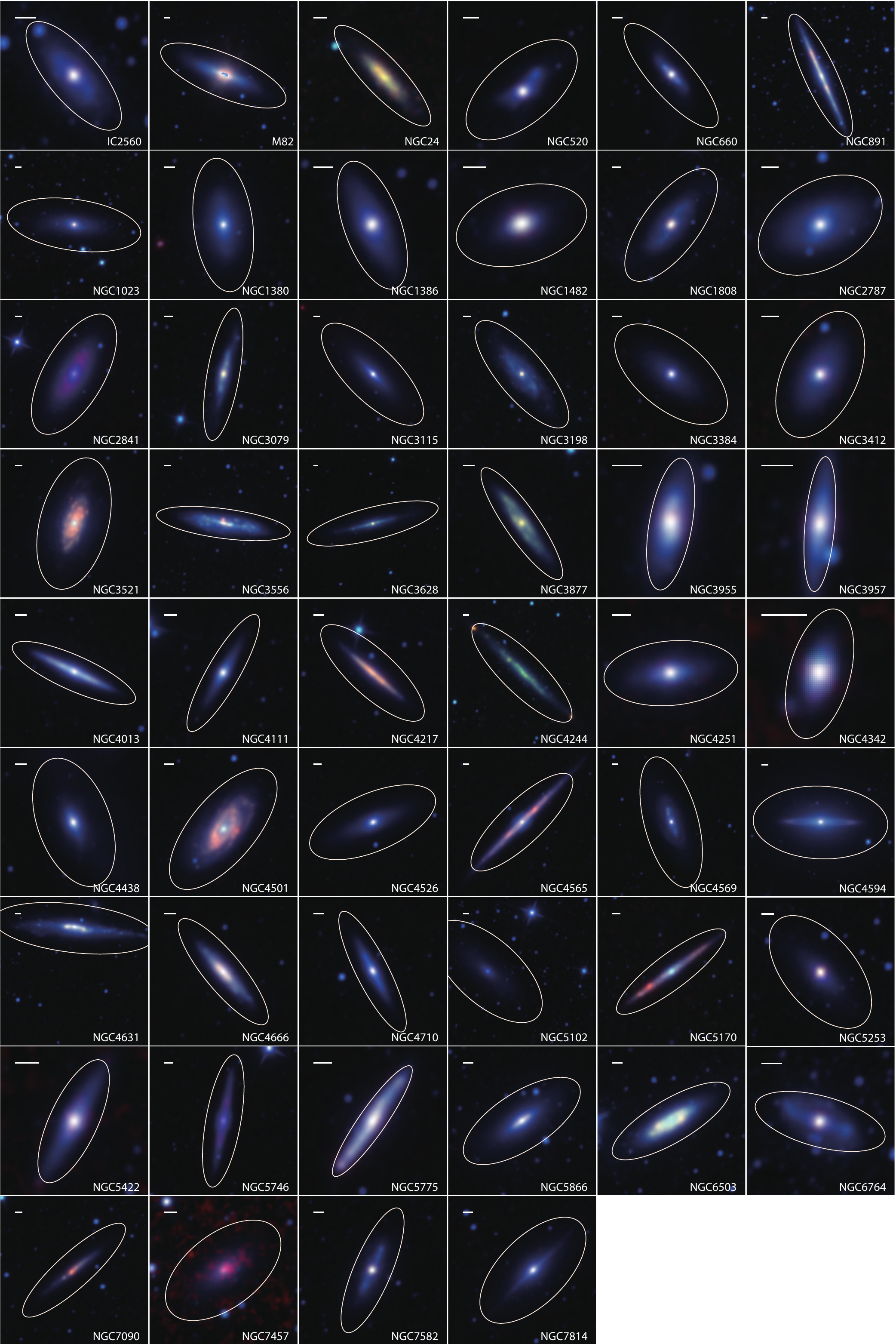}
    \caption{The \wise~montage of our sample galaxies. Each galaxy is shown in a tri-color image in the RGBA (red, green, and blue) color space in which \wise~W1, W2, and W3 flux intensities are represented by red, green, and blue colors, respectively. Galaxies would look bluer with higher star formation rates. The ellipse marking the boundary for aperture photometry is adopted from the \tmass $K_s$-band measurements. The white line at the top-left corner of each plot denotes a scale of 20\arcsec.}
\label{fig:sky}
 \end{figure*}

 \subsection{Sample}
 Our sample consists of 51 nearby, highly inclined disc galaxies that were initially compiled by \citet{Li2013}.
 They selected their sample based on morphological classification (optical morphological type codes of $-3 \gtrsim TC \lesssim 9$), inclination angle ($i \gtrsim 60^{\circ}$), distance ($d < 30$ Mpc), extinction (neutral hydrogen column density $N_H \lesssim 8\times 10^{20}\rm\ cm$), as well as \chandra data availability (See Table 1 in \citealt{Li2013} for details).

 We remove two galaxies, M82 and NGC 4342, from the original sample, because the fluxes in the central region of M82 are saturated in \wise~W3 and W4 bands, and the corona of NGC 4342 are largely affected by its intracluster medium (ICM, \citealt{Bogdan2012}).

Following \citet{Li2013}, we define subsamples based on the morphological type and environment of the galaxies: late or early type; field or cluster environment. We further label 12 galaxies with reported AGNs activities, IC 2560, NGC 660, NGC 1386, NGC 3079, NGC 4388, NGC 4501, NGC 4565, NGC 4569, NGC 4594, NGC 5866, NGC 6764, and NGC 7582 (\citealt{Veron2006,Croston2008}).

\subsection{\wise~data}
In this section, we analyze the \wise~data and derive a few physical parameters of each sample galaxy.
The \wise~four-band images of our sample galaxies have been retrieved from \allwise Release Atlas Images in the \wise~archive\footnote{http://irsa.ipac.caltech.edu/applications/wise}. Figure~\ref{fig:sky} shows the tri-color image of each galaxy in the RGB (red, green, blue) color space.
 We first perform aperture photometry and fit the morphology of the sample galaxies. We then estimate quantities such as stellar mass and star formation rate, and compare our results with previous findings. %These quantities will form the base for our correlation analysis in the next section.

\subsubsection{Apeture photometry}
\begin{figure*}
  \center
  \includegraphics[width=0.9\textwidth]{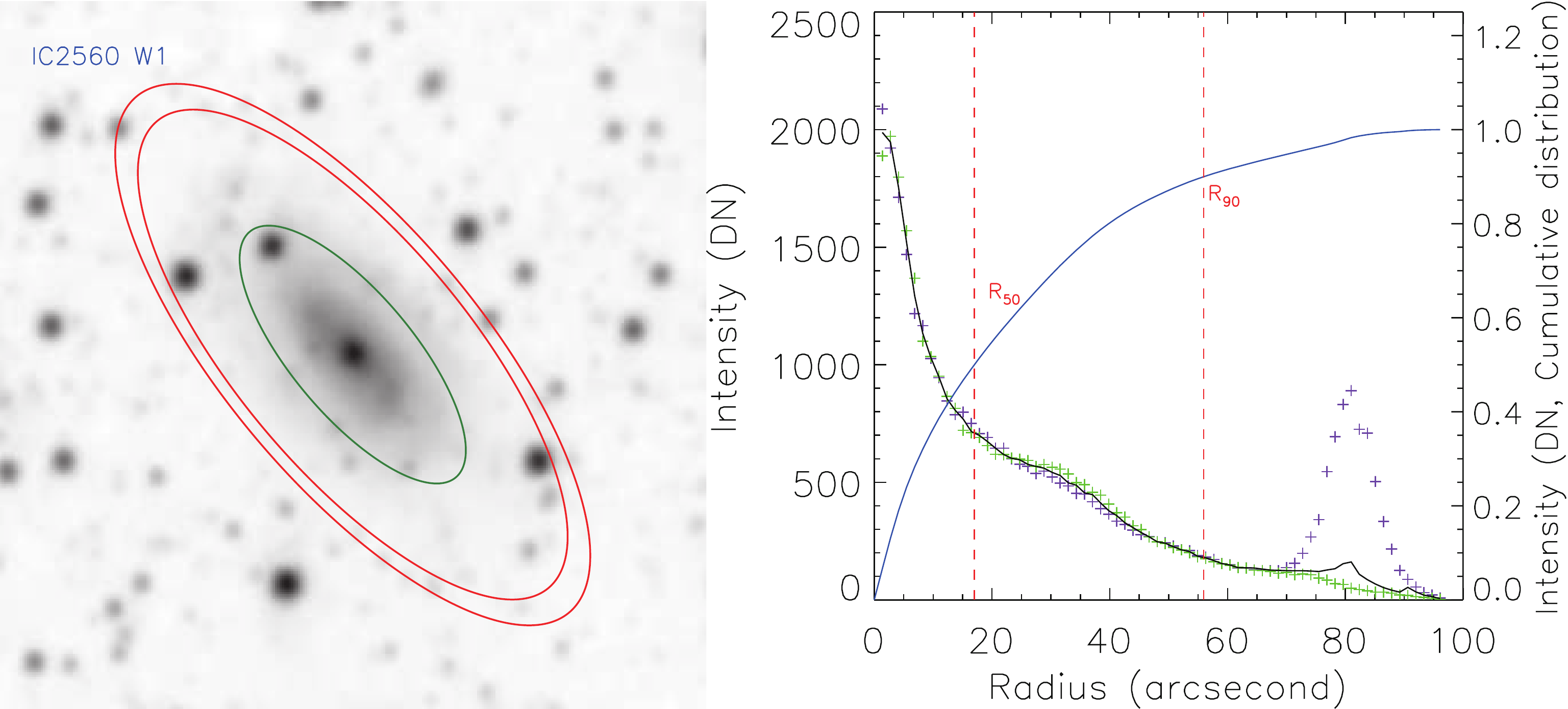}
  \caption{Left panel: \wise~W1 image of IC2560. The green ellipse represents the \tmass $k_s$ band aperture of the galaxy, and the background subtraction region is labeled as the red ring. Right panel: flux intensity (in units of DN\footnote{The unit of intensity is Digital Number (DN), see: http://wise2.ipac.caltech.edu/docs/release/allwise/expsup/sec2\_1a.html}) distribution of the galaxy as the function of the radius along the major axis of the galaxy. Green and purple plus symbols show the distributions along with the two opposite directions of the major axis, and black and blue lines show the median-averaged and cumulative intensity distributions, respectively. Two vertical red dashed lines are labeled at the position of $R_{50}$ and $R_{90}$, respectively.}
  \label{fig:exam}
\end{figure*}

We first define an ellipse for each galaxy as the area (or boundary) of measuring the IR emission profile (see the white ellipse in Figure~\ref{fig:sky}). The major and minor axes of the ellipse are adopted from the \tmass $K_s$ band measurements\footnote{http://ned.ipac.caltech.edu/}. Except for NGC5102, we reduce the $K_s$ aperture by 50\% in the W3 and W4 bands because of the weak emission or the low intensity respect to the background fluctuation in these two bands. To extract the background, we define a concentric elliptical annulus for which the inner and outer major axes equal to 1.9 and 2.1 times of the $K_s$ band major axes, respectively. The background is defined as the median intensity within that annulus. In the left panel of Figure~\ref{fig:exam}, the green line marks the \tmass boundary of the galaxy, while the background subtraction region is labeled with the red lines.

We also adopt a simplified method to calculate the IR emission profile. Specially, we define $r$ as the distance along the major axis to the center, and the total IR intensity is measured within an rectangle region from $r$ to $r+\Delta r$. The vertical boundary of this rectangle region is the ellipse that encompass the entire IR emission (the white ellipse in Figure 1). Then we obtain the 1D intensity distribution along the positive $I_p(r)$ and negative directions $I_n(r)$ of the major axes of the galaxy. Assuming a symmetric distribution of the intrinsic flux of the galaxy, we take a median average the intensity profiles at both directions to obtain an average distribution ($I(r)=median(I_p(r-w), ..., I_p(r+w), I_n(r-w), ..., I_n(r+w))$, where the smooth width $w$ is 6\arcsec). Most of the contributions from point like sources (e.g., the bright source at the top-left corner of the ellipse) have been removed in this median-averaged process. We then obtain the total intensity of the galaxy from the $I(r)$. The blue line shows the cumulative distribution of the intensity.
%To eliminate the point source contamination .
We have also calculated the $R_{50}$ and $R_{90}$ (labeled in red-dashed lines), defined as 50\% and 90\% cumulative intensities, respectively. In Table~\ref{tab:con}, we list both $R_{50}$ and $R_{90}$ for all the four \wise~bands.

Three kinds of corrections, aperture correction, color correction, and calibration correction for W4, were applied for the \wise~co-added images. We refer the readers to Section 3.5 of \citet{Jarrett2013} for details. We have also taken into account the foreground Galactic extinction when deriving the integrated flux. The extinctions for the four \wise~bands in the Milky Way are $A_{3.4\mu m} = 0.056A_{\rm v}$, $A_{4.6\mu m} = 0.049A_{\rm v}$, $A_{12\mu m} = 0.049A_{\rm v}$, $A_{22\mu m} = 0$, respectively (\citealt{Jarrett2013}), where $A_{\rm v}$ was obtained from the  \iras~database \footnote{http://irsa.ipac.caltech.edu/applications/DUST/}. All magnitudes in this paper have been defined in the Vega System. The measured fluxes in the four \wise~bands are listed in column 1--4 of Table~\ref{tab:sfrsm}. %Here the fluxes of galaxy M82 in W3 and W4 are the lower limit because its intensity in the central region is saturated.

\subsubsection{Stellar mass}
We follow \citet{Jarrett2013} to estimate the stellar mass of the sample galaxies. By correlating the stellar mass estimated with \tmass $K_s$ band magnitude and the \wise~($W1-W2$) as well as ($W2-W3$) colors, \citet{Jarrett2013} found the following empirical relations to estimate the mass-to-light ratio:
  \begin{eqnarray}
  \wise \,\, 3.4 \mu m: \,\, log \, (M_*(K_s)/L_{W1}) \,[M_\odot / L_\odot] \, = \, \nonumber \\
-0.246 (\pm 0.027) \, - \, 2.100(\pm0.238) \, (W1-W2) \label{equi:W1sm} \\
  \wise \,\, 3.4 \mu m: \,\, log \, (M_*(K_s)/L_{W1}) \,[M_\odot / L_\odot] \, = \, \nonumber \\
-0.192 (\pm 0.049) \, - \, 0.093(\pm0.016) \, (W2-W3) \label{equi:W2sm}
\end{eqnarray}
Here $L_{W1}$ is the W1 luminosity.% $L_{W1} = 10^{-0.4(M(band)-M_\odot(band))}$, where M(band) and $M_\odot$(band) are the absolute magnitude of the source and solar magnitude measured in this band. The $M_\odot$(band) of \wise 4-band are 3.24, 3.27, 3.23 and 3.25, respectively (\citealt{Oh2008}).

We list the stellar mass calculated with these two empirical formula in columns 6 and 7 of Table~\ref{tab:sfrsm}. In the right panel of Figure~\ref{fig:sfrsm}, we compare the stellar mass estimated with \tmass $K_s$ band magnitude with those estimated with the \wise~data. The stellar mass measured based on the $W2-W3$ color has a much tighter correlation with the $K_s$ band magnitude with less scatter ($r_s=0.969\pm0.001$, $RMS=0.10\pm0.02~dex$).

\subsubsection{Star formation Rate}
\citet{Jarrett2013} suggested that fluxes measured in \wise~W3 and W4 bands strongly correlate with those measured with \emph{Spitzer} 24$\mu m$ fluxes of 17 nearby galaxies. In this work, we focus on the global \sfr~calculated in W3 and W4 bands, adopting the two equations  from \citet{Jarrett2013}:
  \begin{eqnarray}
  \wise~W3: \,\, \sfr_{IR} \, (\pm 0.28) \, (M_\odot \, yr^{-1}) \nonumber \\
   = 4.91 ( \pm 0.39) \times 10^{-10} \, \nu L_{12} (L_\odot)\label{equi:W3SFR} \\
  \wise~W4: \,\,  \sfr_{IR} \, (\pm 0.04) \, (M_\odot \, yr^{-1})  \nonumber \\
   = 7.50 (\pm 0.07) \times 10^{-10} \, \nu L_{22} (L_\odot) \label{equi:W4SFR}
  \end{eqnarray}
Here the spectral luminosity $vL_{v}$ is normalized by $L_\odot$.

  \begin{figure*}
   \center
    \includegraphics[width=1.0\textwidth]{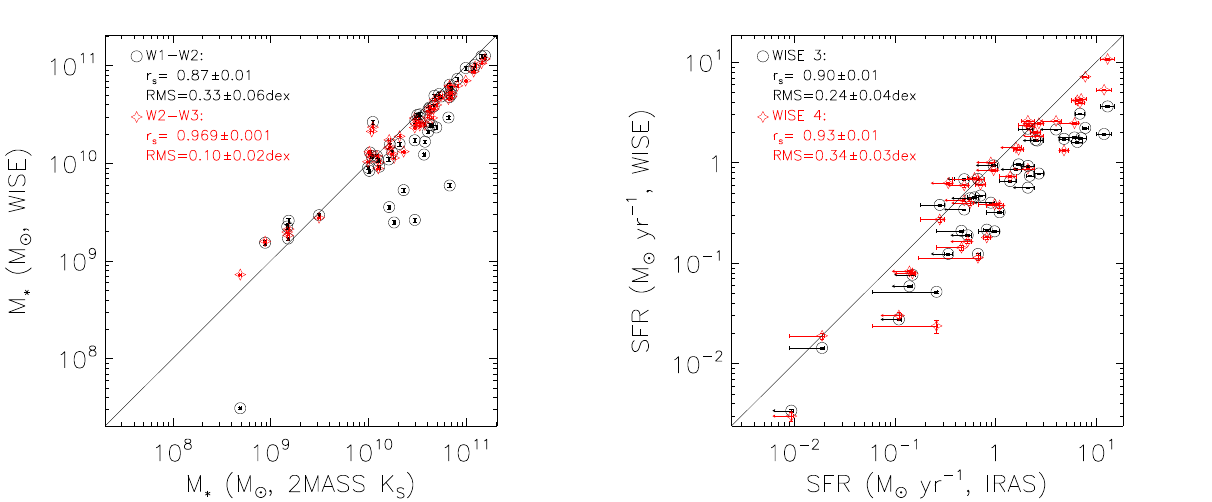}
	  \caption{Left panel: \wise~versus \tmass estimated stellar mass. The black and red symbols represent stellar masses calculated by Eq.~\ref{equi:W1sm} and Eq.~\ref{equi:W2sm}, respectively. Right panel: \wise~versus \iras~estimated \sfrs. The black and red symbols represent W3 and W4-measured \sfrs, respectively. The Spearman's rank and the RMS of each correlation have been labeled on the top left of each panel.}
  \label{fig:sfrsm}
  \end{figure*}

We compare our measurements of the \sfr~with those measured with \iras~for overlapped sample galaxies. The fluxes of part of our galaxies in \iras~were obtained from \citealt{Fullmer1989}. The total IR luminosity is defined as (\citealt{Rice1988}): L$_{IR}$=5.67$\times$10$^{5}$ $d^{2}(13.48 f_{12}+5.16 f_{25}+2.58 f_{60}+ f_{100})L_\odot$ , where $d$ is the distance, and $f_{12}, f_{25}, f_{60}, f_{100}$ are the \iras~fluxes measured at 12, 25, 60, and 100 $\mu m$, respectively. Using the relation of \citealt{Kennicutt1998}, the \iras~\sfrs of our galaxies have been estimated from the $L_{IR}$. In the right panel of Figure~\ref{fig:sfrsm} we show the comparison of the \sfrs measured with the \wise~and \iras~data. The \sfr~estimated with the W4 band luminosity ($r_s=0.93\pm0.01$, $RMS=0.34\pm0.03~dex$) is better consistent with the \iras~\sfr~than the W3 \sfr~($r_s=0.90\pm0.01$, $RMS=0.24\pm0.04~dex$). The large deviation from the best-fit relations is that due to the poorer sensitivity and resolution of \iras~that sometimes only an upper limit on the flux is obtained (see \citealt{Li2013} Table 4).

\subsubsection{Inclination angle}
To accurately describe the morphology of each galaxy, we use the image fitting code, IMFIT, to fit the 2D galaxy image (\citealt{Erwin2015}).  Using this code, the morphology of each galaxy can be decomposed into two components. The first component is a s$\acute{e}$rsic profile (\citealt{Sersic1968}) describing the surface brightness of the bulge:
  \begin{eqnarray}
  & I(r)=I_e exp\left\{-b_n\left[\frac{r}{r_e}^{1/n}-1\right]\right\} \label{equi:sersic}.
  \end{eqnarray}
Here $r_e$ is the half-light radius, $I_e$ is the surface brightness at the radius $r_e$, and $b_n$ is a dimensionless constant to ensure half of the total luminosity is enclosed within $r_e$ (\citealt{Erwin2015}). The second component is an exponential profile for the disc:
  \begin{eqnarray}
  & I(r) = I_0 exp(-r/h) \label{equi:expo}.
  \end{eqnarray}
Here $I_0$ and $h$ are the central surface brightness and the exponential scale length, respectively.

To fit the morphology of each galaxy, we use the IDL code MEANCLIP (\citealt{Landsman1993}) to obtain the sky background and the corresponding 1$\sigma$ fluctuation. We also set the initial values of the position angle and ellipticity to those obtained from the \tmass catalogue. We made use of the uncertainty map of each galaxy retrieved from the \wise~archive as the weight map.% Finally, we use $\chi^2$ statistics for minimization.

We calculated the inclination angle of the galaxies $i$ from the fitted disk ellipticity ($\cos(i)\equiv 1-ell$) using W1 and W2 images. We adopt the average value of the measurements in these two bands as the final inclination angle of the galaxies (see Column 8 in Table~\ref{tab:con}).

\subsection{X-ray data}
\label{sec:mor}
The X-ray data analysis is detailed in \citet{Li2013}. We only present a brief summary here. After standard data calibration, the point source removed, exposure corrected Chandra images were used to calculate the coronal X-ray luminosity and to extract the vertical (perpendicular to the disk plane) and horizontal (along with the disk plane) X-ray surface brightness profile. %The 0.5-2 keV X-ray luminosity was estimated using a pure 1-T plasma, XSPEC model Mewe-Kaastra-Liedahl (MEKAL, \citealt{Mewe1985}), when the photon counting statistics is good; or assuming a thermal plasma of a typical 0.3 keV temperature for the corona with low photon counts.

The vertical profiles are extracted in a region where $|r|<D_{25}/4$. Here $r$ is the radial distance along the galactic disk from the nucleus of the galaxy and $D_{25}$ is optical diameter. In this procedure, we exclude a significant valley (only occurring on the near side of the disk) of profile along the galactic disc (Figure 5 in \citealt{Li2013}). Then the total luminosity of the extraplanar emission is corrected for the loss of the filtered-out ``disc" range (Section 5.1 in \citealt{Li2013}). By doing this, we have excluded the effect of the ISM absorption from the calculation of scale heights. Using this profile, we define several vertical heights of the extraplanar emission, $H_{50}$, $H_{75}$, and $H_{95}$, enclosing 50\%, 75\%, and 95\% of the total flux, respectively. The vertical scales are calculated by averaging the positive and negative sides of galaxies after removing the obscured disk (Table~\ref{tab:xray}). All the vertical heights in this paper are corrected by the inclination angles--dividing $\sin(i)$, where $i$ is the inclination angle. To test the dependence of inclination, we generate a random number ranged from 60 to 90 for each galaxy as the inclination angle. Then we corrected the scale heights by the random inclination angle, and redo the relations in Section 3. We find the inclination correction hardly affect correlations of all the relations, and the uncertainty is well within the statistical and systematic errors.

\begin{figure}
\center
    \includegraphics[width=0.50\textwidth]{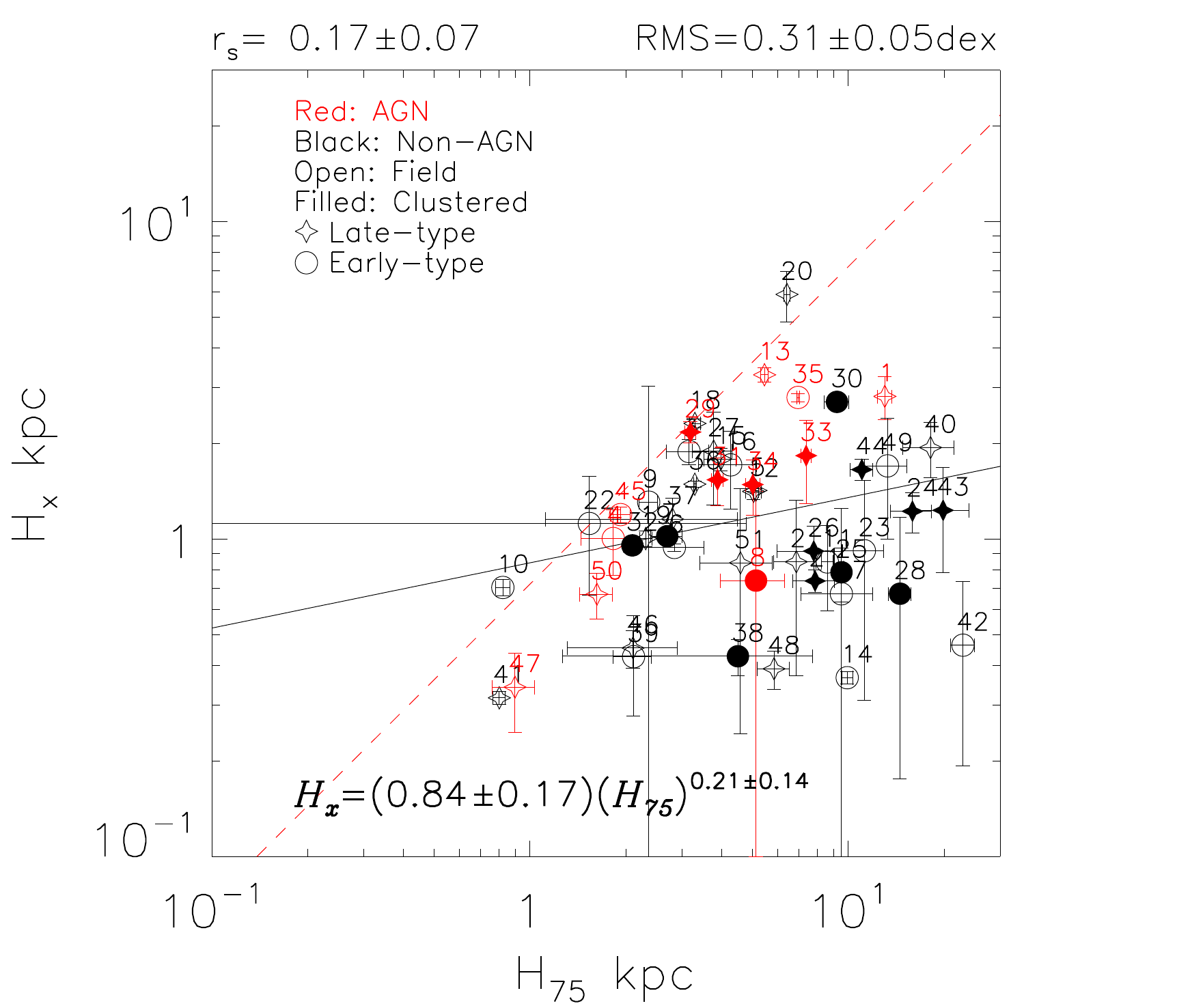}
 \caption{Comparison between the characteristic exponential scales and the model-independent scale $H_{75}$. The red dotted line represents $H_{x} = H_{75}/ln4$ for a single exponential model. The red or black symbols show the galaxies with or without AGN, respectively. The open and filled symbols show the field galaxies and cluster galaxies, respectively. The star and circle symbols show the interacting and isolated galaxies, respectively.}
 \label{fig:hhx}
\end{figure}

In addition to the morphological parameters $H_{50}$, $H_{75}$ and $H_{95}$, we also obtain the exponential scale height of the galaxies in the vertical direction ($H_X$) from \citealt{Li2013}. $H_X$ is calculated on both sides, above and below the galactic disk, and the averaged values are used for the statistical analysis below. We compare $H_X$ and $H_{75}$ in Figure~\ref{fig:hhx}. Assuming an exponential soft X-ray vertical intensity profile, $H_X$ is equivalent to $H_{75}$/ln4, which is shown as the red dotted line in the figure. Clearly for some galaxies $H_X$ is far less than $H_{75}$. In many of these galaxies, there is a strong X-ray emitting close to the disk dominating the extraplanar emission, so a single exponential model may not well describe the vertical distribution of the soft X-ray emission. We will, therefore, adopt the model-independent ``scale heights" directly obtained from observations ($H_{75}$ and $H_{95}$) in the following statistical analysis.

\section{Analysis and Result}
\label{cor}
We use Spearman's rank order coefficient ($r_s$; ranging from -1 to 1) to evaluate the goodness of all the correlations in this work. A correlation coefficient of $0\leqslant |r_s| \leqslant 0.3$, $0.3 < |r_s| \leqslant 0.6$, or $|r_s| > 0.6$ is considered as no, weak, or strong correlation, respectively. We describe the correlation as a linear relation in logarithmic scale and calculate the root mean square (RMS) of the data points around the best-fit relation. The 1$\sigma$ errors quoted in the following analysis are estimated by bootstrap-with-replacement data (1000 times).
We adopt the stellar mass calculated with Equation~\ref{equi:W2sm} and \sfr~calculated with Equation~\ref{equi:W4SFR}. 
We also define a specific parameter of the galaxies: the SF intensity $I_\sfr = \sfr/(\pi R^2_{W4})$, where $R_{W4}$ are $R_{90}$ in W4 band.
%We define two scale heights, innermost scale $H_{75}$ which trace the boundary of observed X-ray morphology, and outermost scales $H_{95}$ which suppose to closely relate to the SF feedback.

\begin{figure*}
\center
    \includegraphics[width=1.00\textwidth]{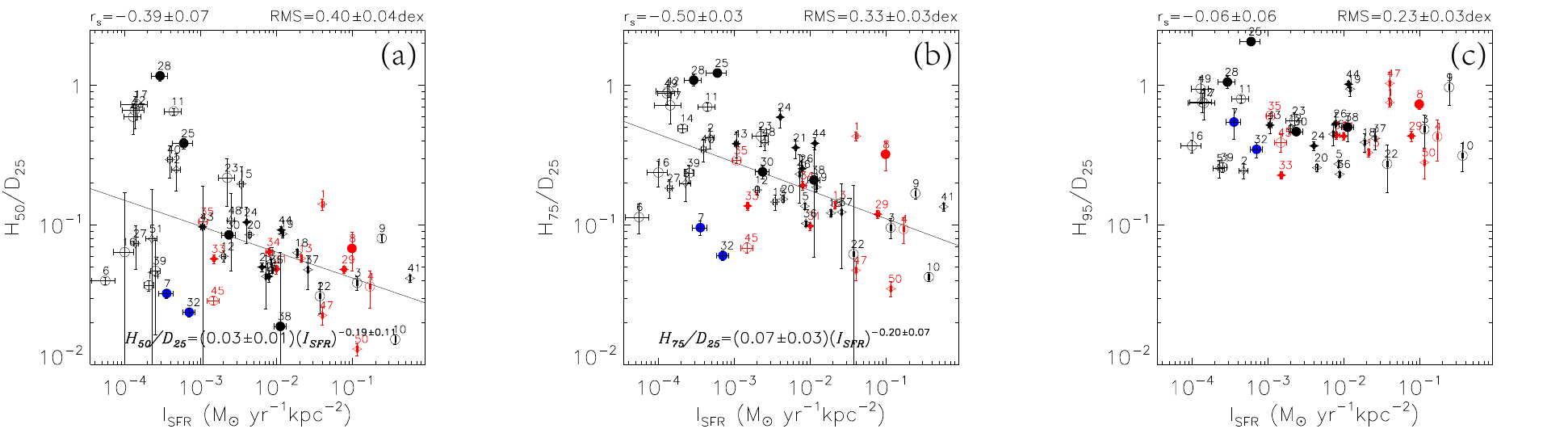}
	\caption{Comparison of the $H_{50}/D_{25}$ (a), $H_{75}/D_{25}$ (b), and $H_{95}/D_{25}$ (c) with the SF intensity of the host galaxies. The galaxies labeled blue symbol are potentially contaminated by nearby extended X-ray source. The other symbols are the same as those in Figure~\ref{fig:hhx}.}
 \label{fig:H}
\end{figure*}

%why normalized D25, why compare to ISFR
%To remove the uncertainty of galaxy distance, we first compare the normalized vertical extension of the diffuse soft X-ray emission ($H_{50}/D_{25}$, $H_{75}/D_{25}$, and $H_{95}/D_{25}$) to the SF intensity ($I_\sfr$).
We first compare the vertical extension of the diffuse soft X-ray emission to the SF intensity ($I_\sfr$). To remove the uncertainty of galaxy distance, we normalize the extension by the $D_{25}$ ($H_{50}/D_{25}$, $H_{75}/D_{25}$, and $H_{95}/D_{25}$) which derive from the third-party observations.
Figure~\ref{fig:H} shows different trends on the $H_{50}/D_{25}$ or $H_{75}/D_{25}-I_\sfr$ and $H_{95}/D_{25}-I_\sfr$ relations. The former, $H_{75}/D_{25}-I_{\sfr}$ relation for example, shows a moderate negative correlation with $r_s=-0.50\pm0.03$ and $RMS=0.33\pm0.03~dex$, while the latter shows no correlation with $r_s=-0.07\pm0.06$ and $RMS=0.23\pm0.03~dex$.
For the $H_{75}/D_{25}-I_\sfr$ relation, the correlation with field galaxies becomes tighter ($r_s=-0.61\pm0.04$, $\Gamma=-0.22\pm0.08$).
 There is no significant difference on the tightness of correlation among the other sub-samples.
%And the negative correlations are not changed for galaxies without AGN ($r_s = -0.44\pm0.04$, $\Gamma=-0.21\pm0.07$) and for ones without interacting or meger ($r_s=-0.47\pm0.03$, $\Gamma=-0.21\pm0.06$).
% do not more mention AGN
\begin{figure*}
   \center
    \includegraphics[width=0.90\textwidth]{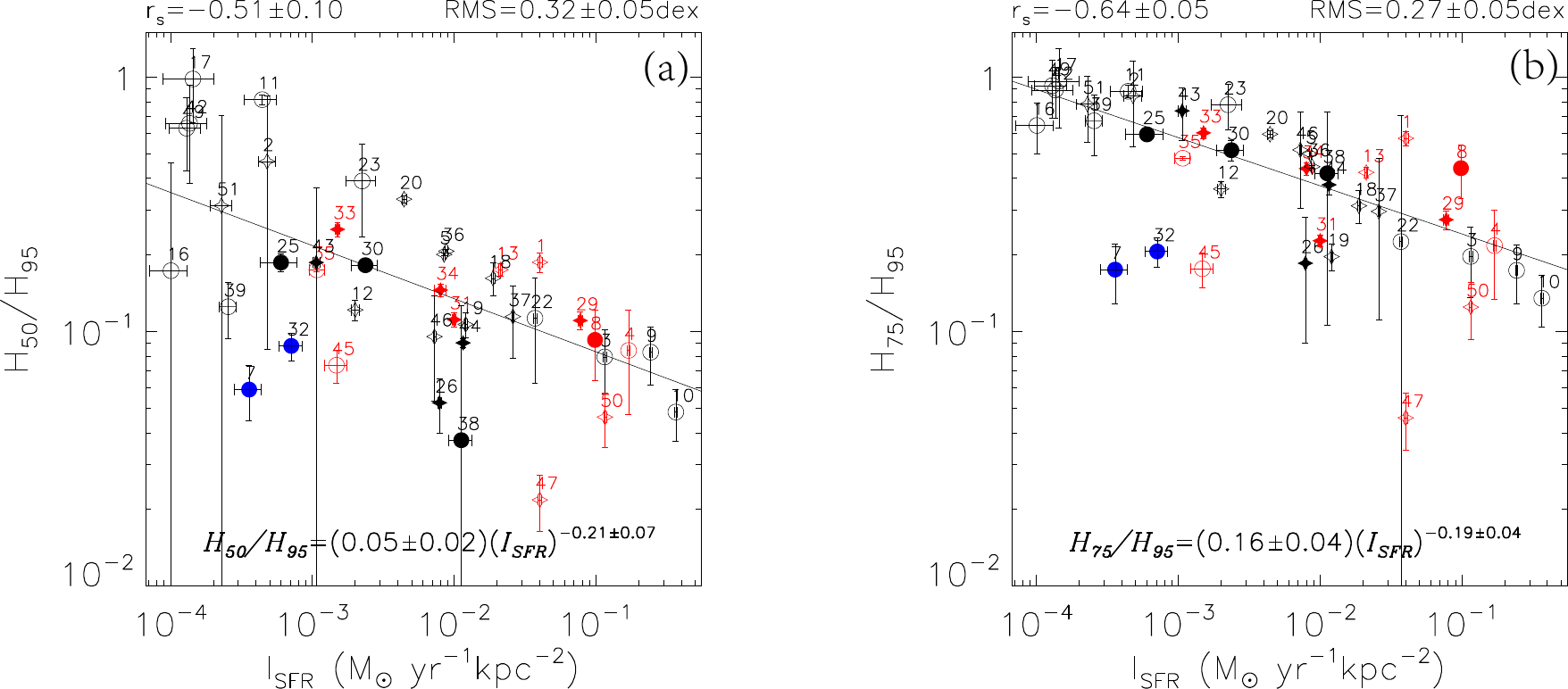}
	\caption{Comparison of the X-ray vertical concentration $H_{50}/H_{95}$ (a) and $H_{75}/H_{95}$ (b) with the SF intensity. The symbols are the same meaning as those in Figure~\ref{fig:H}.}
    \label{fig:con}
\end{figure*}

%However, the scale $H_{95}/D_{25}$ show no correlations with the SF intensity, indicating that the dominated component may different.
To further quantify the spatial extent of the diffuse X-ray emission, we define X-ray concentration parameters $H_{50}/H_{95}$ and $H_{75}/H_{95}$ along the vertical direction (Figure~\ref{fig:con}).
Since the $H_{95}$ of six galaxies (NGC 24, NGC 3556, NGC 4217, NGC 4526, NGC 5102, and NGC 5422) is only accessible in one side, the $H_{75}$ related concentration for example, is calculated by $H_{75+}/H_{95+}$ or $H_{75-}/H_{95-}$ instead of $H_{75}/H_{95}$.
%Because of the background fluctuation of the outermost region of galaxies, we fail to measure the $H_{95}$ on one or two sides of some galaxies (``---" in Table~\ref{tab:xray}). For the only one side measured $H_{95}$, the concentrations are calculated by the corresponding side of $H_{75}$.
We obtain a tighter anti-correlation of $H_{75}/H_{95}$ with $I_{\sfr}$ ($r_s = 0.63\pm0.04$, $\Gamma = -0.18\pm0.04$) than the $H_{75}/D_{25}$. The anti-correlation for field galaxies ($r_s = -0.76\pm0.05$, $\Gamma = -0.21\pm0.04$) is also much stronger. %The galaxies with AGN, interacting or merger galaxies would not affect this concentration correlation ($r_s = -0.60\pm0.06$, $\Gamma = -0.19\pm0.05$ for galaxies without AGN; $r_s = -0.67\pm0.08$, $\Gamma = -0.20\pm0.04$ for isolated galaxies).

%\begin{figure}
%\center
%    \includegraphics[width=0.80\textwidth]{./pic/Isfr_vs_H}
%    \includegraphics[width=0.49\textwidth]{./pic/l_Isfr_vs_h75d25}
   %\includegraphics[width=0.49\textwidth]{./pic/l_Isfr_vs_h75h95}
	%\caption{Comparison of the concentration $H_{75}/H_{95}$ with the SF intensity for the low-mass galaxies ($M_*<3\times10^{10}M_\odot$).}
 %\label{fig:lH}
%\end{figure}

Furthermore, we notice that the $H_{95}$ values of two galaxies, NGC 1380 and NGC 4526 (No. 7 and 32) are overestimated because the X-ray emission of each is contaminated by another diffuse X-ray source perpendicular to the galaxy disc (see Figure 4 in \citealt{Li2013}). The $H_{75}/H_{95}-I_{\sfr}$ anti-concentration becomes even tighter with $r_s=-0.73\pm0.05$ and $RMS=0.25\pm0.06~dex$ after removing these two outlier galaxies.

\section{Discussion}

Compared with previous statistical studies of the X-ray luminosity-related relations (e.g. \citealt{Li2013a,Strickland2004,Tuellmann2006,Wang2016}), the vertical heights not only provide additional spatial information, but also help distinguish the contribution of different components. 
As the outermost soft X-ray vertical extension of the extraplanar emission observed in our sample ($H_{95}$) is much smaller than the typical cooling radius of galaxies (\citealt{Bregman2018}), it is likely $H_{95}$ is affected more by stellar feedback than accretion from the IGM, we therefore focus on stellar feedback in the following discussion.

We suggest that the two clearly different trends in the $H_{50}$ or $H_{75}$ versus $I_{\sfr}$ (Figure~\ref{fig:H}a,b) and $H_{95}$ versus $I_\sfr$ plots (Figure~\ref{fig:H}c) are caused by the contributions of the two significantly different soft X-ray emitting components. One component is the optically thin thermal emission from the hot plasma itself. The other is the emission generated at its interface with cool gas. Interface processes such as CXE, cooling/condensation and/or turbulent mixing of different gas phases, can enhance the soft X-ray emission (e.g. \citealt{Liu2011,Li2011}). The thermal plasma emission should be extended and radiatively inefficient due to the low density of the gas (\citealt{Strickland2007,Strickland2009}). On the other hand, the cool-hot gas interaction is largely affected by the spatial distribution of the cool gas, which often appears to be much more confined than the hot gas around nearby galaxies. 
Therefore, it is likely that $H_{50}$ (or $H_{75}$) and $H_{95}$ trace the cool-hot interaction and the extended diffuse hot gas, respectively.

Although the total X-ray emission of a galaxy seems positively dependent on the SF activity (characterized with the total \sfr, \citealt{Strickland2004,Grimes2005,Tuellmann2006,Li2008,Li2013a}), the efficiency converting the stellar feedback energy to X-ray emission is actually negatively dependent on the SF intensity (\citealt{Li2013a,Wang2016}).
A larger fraction of stellar feedback energy is carried out by the galactic wind when the SF intensity increases. This component has low density thus inefficient in radiating X-ray. On the other hand, most of the enhanced X-ray emission could be attributed to the cool-hot gas interaction.
It is natural to assume that the strength of cool-hot gas interaction depend on the amount of cool gas and how well it mixes with the hot gas.
\citet{Jo2018} reports the positive correlations between the SF intensity and the scale height of extraplanar H$\alpha$ and UV with 38 nearby edge-on, late-type galaxies. %They suggested that the SF intensity strongly correlates with the scale height of UV and H$\alpha$ emission. 
They suggested that the extraplanar dust and cool gas traced by UV and H$\alpha$, respectively, become thicker with the higher SF intensity. 
This projected envelope of cool gas most likely represents an inter-mixing with the extended hot corona and enhances X-ray emission via cool-hot gas interaction.

The vertical extension of cool gas ($\sim 1~kpc$, \citealt{Mccormick2013,Jo2018}) is typically much smaller than $H_{50}$ of the hot gas in our sample. Therefore, with increasing SF intensity, the vertical distribution of X-ray emission, strongly enhanced by cool-hot gas interaction close to the disk, becomes significantly steeper, resulting in smaller $H_{50}$ and $H_{75}$ (Figure~\ref{fig:H}a,b), as well as lower $H_{50}/H_{95}$ and $H_{75}/H_{95}$ ratios (Figure~\ref{fig:con}a,b). %The larger heights $H_{75}$ (0.8--22.9 kpc) which enclosed most of the cool gas is more representative than the $H_{50}$. Indeed, the correlations of $H_{75}$ mentioned in \S\ref{cor} systematically present better than the ones of $H_{50}$, at least in our sample.
On the other hand, $H_{95}$ shows clearly different dependence on $I_{\sfr}$ from $H_{50}$ and $H_{75}$, with no significant correlation (Figure~\ref{fig:H}c). This difference apparently indicates that $H_{95}$ traces another X-ray emitting component which is extended but not directly related to cool-hot gas interaction. We speculate that this component is the extended thermal plasma which is likely blew out by the large scale galactic outflow. The outflow could be very tenuous and extended (up to $\sim60~kpc$, e.g. \citealt{Strickland2000,Tang2009,Tripp2011}) so the measured size of it in $H_{95}$ is determined by the sensitivity of the observations, instead of its intrinsic properties.

The diffuse X-ray morphology of galaxies may be further affected by the presence of AGN and the ICM. The presence of AGN may further increase the scatter of the relations. We mark all the galaxies with identified AGN with red-color symbols to show such scatters. 
The X-ray morphology of clustered galaxies may be highly affected by the presence of the dense ICM. The anti-correlations of field galaxies shown in Figure~\ref{fig:H} and Figure~\ref{fig:con} appear to be much tighter than those of the clustered ones (see Section~\ref{cor} for details), which confirms that the relations between the X-ray morphology of the extraplanar emission and the SF intensity found in this paper is indeed due to internal instead of environmental processes.

\section{summary}

X-ray emission of the hot galactic corona provides rich information on processes that impact galaxy formation and evolution. In this work, we study a sample of 51 nearby, edge-on, disk galaxies. Specifically, we focus on the correlations between the diffuse soft X-ray morphology obtained from \chandra, and the SF intensity obtained from \wise. Our main results are summarized as follows:

(1) We calculate the \sfr~and the $M_*$ of sample galaxies from \wise~galaxy images and obtain the characteristic radius $R_{50}$ and $R_{90}$ of galaxies in \wise~four bands. We also calculate the inclination angle of each galaxy by fitting the ellipticity of its stellar disk.  We further characterize the vertical extension of the extraplanar emission with a few parameters ($H_{50}$, $H_{75}$, and $H_{95}$) defined with the X-ray intensity profile of each galaxy.

(2) We find that the single exponential model brings in significant bias in describing the vertical distribution of the X-ray emission of our sample galaxy. 
The vertical extension of the extraplanar emission characterized with $H_{50}$ or $H_{75}$ and the vertical concentration characterized with $H_{50}/H_{95}$ or $H_{75}/H_{95}$ are both tightly anti-correlated with the SF intensity. However, we do not find any correlation between $H_{95}$ and the SF intensity.

(3) We also find tighter correlations of $H_{75}/D_{25}-I_\sfr$ and $H_{75}/H_{95}-I_\sfr$ relations for field galaxies, and no significant difference in the tightness of correlation among the other sub-samples.

Our findings suggest that there are likely two components for a typical corona: the concentrated corona mainly produced by cool/hot gas interaction, and a more extended corona which should contain a significant fraction of energy and mass but disproportionally low X-ray emission. %The implication is that we cannot use the bulk soft X-ray emission to represent the hot coronal properties, as the extended component with weak X-ray emission may be indeed more important.

\section*{Acknowledgements}
This work is supported by the National Key R\&D Program of China No. 2017YFA0402600, the Fundamental Research Funds for the Central Universities at Xiamen University under grant No. 20720190115, and NSFC grants No. 11525312, 11333004, 11443009, U1531130, and 11333004. T.F. is also supported by the Specialized Research Fund for the Doctoral Program of Higher Education (SRFDP; No. 20130121110009).

This publication makes use of data products from the Wide-field Infrared Survey Explorer, which is a joint project of the University of California, Los Angeles, and the Jet Propulsion Laboratory/California Institute of Technology, funded by the National Aeronautics and Space Administration.

\begin{deluxetable*}{lccccccccc}
\centering
\tiny
  \tabletypesize{\tiny}
  \tablecaption{Characteristic radius of the sample galaxies}
  \tablewidth{0pt}
  \tablehead{
	  \colhead{Number} & \colhead{Name} & \colhead{$R_{50}^{1}$} & \colhead{$R_{50}^{2}$} & \colhead{$R_{50}^{3}$} & \colhead{$R_{50}^{4}$} & \colhead{$R_{90}^1$} & \colhead{$R_{90}^2$} & \colhead{$R_{90}^3$} & \colhead{$R_{90}^4$}  \\%& \colhead{$f_{W1}$} & \colhead{$f_{W2}$} & \colhead{$f_{W3}$} & \colhead{$f_{W4}$} \\
	& & (arcsec) & (arcsec) & (arcsec) & (arcsec) & (arcsec) & (arcsec) & (arcsec) & (arcsec) \\%& (Jy) & (Jy) & (Jy) & (Jy) \\
 & (1) & (2) & (3) & (4) & (5) & (6) & (7) & (8) & (9) \\%& (9) & (10) & (11) & (12)
}
\startdata
1   &  IC2560 & $ 14.8\pm  0.1$ & $ 12.7\pm  0.1$ & $  9.0\pm  0.1$ & $  9.2\pm  0.1$ & $  57.8\pm   0.1$ & $  52.6\pm   0.2$ & $  45.8\pm   0.2$ & $  32.0\pm   0.5$  \\
2   &  NGC24 & $ 28.3\pm  0.1$ & $ 30.1\pm  0.2$ & $ 29.0\pm  0.6$ & $ 34.6\pm  2.7$ & $  94.3\pm   0.2$ & $  95.1\pm   0.7$ & $  77.6\pm   2.3$ & $ 101.5\pm   6.1$   \\
3   &  NGC520 & $ 13.4\pm  0.1$ & $ 10.1\pm  0.1$ & $  7.4\pm  0.1$ & $  9.0\pm  0.1$ & $  51.5\pm   0.1$ & $  45.1\pm   0.2$ & $  35.4\pm   0.2$ & $  28.4\pm   0.3$  \\
4   &  NGC660 & $ 12.1\pm  0.1$ & $  7.1\pm  0.1$ & $ 12.3\pm  0.1$ & $ 11.0\pm  0.1$ & $  57.3\pm   0.1$ & $  38.5\pm   0.2$ & $  52.7\pm   0.1$ & $  39.8\pm   0.2$  \\
5   &  NGC891 & $ 74.1\pm  0.2$ & $ 73.7\pm  0.1$ & $ 87.4\pm  0.2$ & $ 81.8\pm  0.3$ & $ 198.0\pm   0.1$ & $ 192.8\pm   0.1$ & $ 188.2\pm   0.2$ & $ 179.2\pm   0.7$  \\
6   &  NGC1023 & $ 26.4\pm  0.1$ & $ 27.0\pm  0.1$ & $ 34.1\pm  1.5$ & $ 68.6\pm 13.2$ & $ 143.0\pm   0.2$ & $ 142.4\pm   0.4$ & $ 173.7\pm   7.3$ & $ 240.4\pm  11.4$ \\
7   &  NGC1380 & $ 16.5\pm  0.1$ & $ 17.1\pm  0.1$ & $ 15.1\pm  0.2$ & $ 17.2\pm  1.2$ & $  76.6\pm   0.1$ & $  77.0\pm   0.1$ & $  75.7\pm   1.5$ & $ 109.5\pm  11.2$ \\
8   &  NGC1386 & $  9.4\pm  0.1$ & $  6.5\pm  0.1$ & $  6.0\pm  0.1$ & $  8.4\pm  0.1$ & $  40.0\pm   1.1$ & $  37.6\pm   0.1$ & $  21.0\pm   0.1$ & $  24.3\pm   0.1$ \\
9   &  NGC1482 & $  5.7\pm  0.1$ & $  5.6\pm  0.1$ & $  6.6\pm  0.1$ & $  8.7\pm  0.1$ & $  22.7\pm   0.1$ & $  21.7\pm   0.1$ & $  20.6\pm   0.1$ & $  24.5\pm   0.1$ \\
10  &  NGC1808 & $ 13.8\pm  0.1$ & $ 11.3\pm  0.1$ & $  9.7\pm  0.1$ & $ 10.3\pm  0.1$ & $  93.0\pm   0.1$ & $  85.1\pm   0.1$ & $  67.1\pm   0.1$ & $  41.8\pm   0.1$ \\
11  &  NGC2787 & $ 11.8\pm  0.1$ & $ 12.8\pm  0.1$ & $ 10.6\pm  0.2$ & $ 12.9\pm  1.7$ & $  56.4\pm   0.1$ & $  56.1\pm   0.1$ & $  52.0\pm   1.7$ & $  65.4\pm   6.8$ \\
12  &  NGC2841 & $ 34.8\pm  0.1$ & $ 37.0\pm  0.1$ & $ 68.4\pm  0.4$ & $ 65.1\pm  1.4$ & $ 139.1\pm   0.1$ & $ 139.5\pm   0.3$ & $ 144.2\pm   1.2$ & $ 142.0\pm   4.1$ \\
13  &  NGC3079 & $ 21.2\pm  0.1$ & $ 18.8\pm  0.1$ & $ 19.9\pm  0.1$ & $ 23.5\pm  0.1$ & $  99.7\pm   0.2$ & $  90.6\pm   0.2$ & $  77.9\pm   0.2$ & $  76.1\pm   0.6$ \\
14  &  NGC3115 & $ 17.8\pm  0.1$ & $ 17.5\pm  0.1$ & $ 21.2\pm  1.1$ & $ 84.5\pm 12.6$ & $ 114.6\pm   0.2$ & $ 111.5\pm   0.5$ & $ 136.8\pm   9.3$ & $ 239.6\pm   8.2$ \\
15  &  NGC3198 & $ 40.2\pm  0.1$ & $ 38.9\pm  0.2$ & $ 37.8\pm  0.3$ & $ 30.4\pm  0.8$ & $ 126.1\pm   0.2$ & $ 120.2\pm   0.6$ & $ 112.4\pm   1.1$ & $ 114.4\pm   3.5$ \\
16  &  NGC3384 & $ 13.5\pm  0.1$ & $ 14.1\pm  0.1$ & $ 14.4\pm  0.9$ & $ 55.2\pm 10.8$ & $  90.2\pm   0.1$ & $  88.7\pm   0.4$ & $ 109.5\pm   7.2$ & $ 169.7\pm   7.3$ \\
17  &  NGC3412 & $ 12.0\pm  0.1$ & $ 13.5\pm  0.1$ & $ 15.3\pm  1.0$ & $ 24.1\pm  6.6$ & $  59.2\pm   0.1$ & $  58.8\pm   0.2$ & $  65.5\pm   3.0$ & $  88.9\pm   9.7$ \\
18  &  NGC3521 & $ 30.0\pm  0.1$ & $ 30.8\pm  0.1$ & $ 50.4\pm  0.1$ & $ 46.0\pm  0.2$ & $ 118.4\pm   0.1$ & $ 117.3\pm   0.2$ & $ 126.7\pm   0.4$ & $ 116.0\pm   1.1$ \\
19  &  NGC3556 & $ 61.7\pm  0.1$ & $ 60.0\pm  0.2$ & $ 55.6\pm  0.1$ & $ 30.7\pm  0.2$ & $ 141.0\pm   0.1$ & $ 138.4\pm   0.2$ & $ 127.6\pm   0.2$ & $ 117.6\pm   0.5$ \\
20  &  NGC3628 & $ 67.2\pm  0.2$ & $ 62.3\pm  0.2$ & $ 56.5\pm  0.3$ & $ 30.4\pm  0.4$ & $ 217.7\pm   0.2$ & $ 211.1\pm   0.5$ & $ 192.2\pm   1.0$ & $ 153.2\pm   2.4$ \\
21  &  NGC3877 & $ 26.7\pm  0.1$ & $ 26.6\pm  0.1$ & $ 31.9\pm  0.1$ & $ 28.1\pm  0.2$ & $  89.3\pm   0.1$ & $  87.1\pm   0.1$ & $  83.7\pm   0.2$ & $  79.9\pm   0.5$ \\
22  &  NGC3955 & $ 11.3\pm  0.1$ & $ 11.5\pm  0.1$ & $ 10.0\pm  0.1$ & $ 11.0\pm  0.1$ & $  31.6\pm   0.1$ & $  33.9\pm   0.1$ & $  25.3\pm   0.1$ & $  27.3\pm   0.3$ \\
23  &  NGC3957 & $ 11.5\pm  0.1$ & $ 12.1\pm  0.1$ & $  9.5\pm  0.1$ & $ 10.2\pm  0.6$ & $  31.8\pm   0.1$ & $  34.1\pm   0.1$ & $  25.7\pm   0.5$ & $  30.0\pm   3.4$ \\
24  &  NGC4013 & $ 23.4\pm  0.1$ & $ 23.8\pm  0.1$ & $ 36.4\pm  0.1$ & $ 35.0\pm  0.4$ & $  80.2\pm   0.1$ & $  79.3\pm   0.1$ & $  85.3\pm   0.3$ & $  82.5\pm   0.9$ \\
25  &  NGC4111 & $  8.2\pm  0.1$ & $  8.6\pm  0.1$ & $  8.8\pm  0.2$ & $ 12.3\pm  0.8$ & $  52.8\pm   0.1$ & $  53.0\pm   0.2$ & $  47.5\pm   2.9$ & $  82.4\pm  11.8$ \\
26  &  NGC4217 & $ 31.8\pm  0.1$ & $ 31.8\pm  0.1$ & $ 32.2\pm  0.1$ & $ 34.8\pm  0.3$ & $ 105.0\pm   0.1$ & $ 102.3\pm   0.2$ & $  86.8\pm   0.2$ & $  83.1\pm   0.7$ \\
27  &  NGC4244 & $ 89.2\pm  0.2$ & $ 88.3\pm  0.5$ & $ 92.9\pm  2.2$ & $128.0\pm  9.0$ & $ 225.0\pm   0.4$ & $ 223.0\pm   1.4$ & $ 269.7\pm   6.6$ & $ 309.6\pm   1.3$ \\
28  &  NGC4251 & $  9.6\pm  0.1$ & $ 10.4\pm  0.1$ & $ 11.9\pm  0.4$ & $ 21.2\pm  4.3$ & $  38.4\pm   0.2$ & $  43.8\pm   0.2$ & $  51.0\pm   2.3$ & $  69.0\pm   5.9$ \\
29  &  NGC4388 & $ 17.9\pm  0.1$ & $  9.3\pm  0.1$ & $ 10.2\pm  0.1$ & $ 10.4\pm  0.1$ & $  84.2\pm   0.2$ & $  71.8\pm   0.5$ & $  41.6\pm   0.3$ & $  37.2\pm   0.7$ \\
30  &  NGC4438 & $ 13.5\pm  0.1$ & $ 13.4\pm  0.1$ & $ 13.0\pm  0.2$ & $ 13.4\pm  0.6$ & $  78.7\pm   0.1$ & $  74.9\pm   0.3$ & $  64.9\pm   1.7$ & $  67.1\pm   7.0$ \\
31  &  NGC4501 & $ 30.6\pm  0.1$ & $ 31.8\pm  0.1$ & $ 38.2\pm  0.1$ & $ 41.1\pm  0.4$ & $ 110.8\pm   0.1$ & $ 110.5\pm   0.2$ & $ 104.9\pm   0.3$ & $ 100.7\pm   1.1$ \\
32  &  NGC4526 & $ 18.9\pm  0.1$ & $ 19.0\pm  0.1$ & $ 11.5\pm  0.3$ & $ 16.3\pm  1.2$ & $ 111.6\pm   0.1$ & $ 112.9\pm   0.5$ & $  90.8\pm   5.1$ & $ 155.2\pm  13.5$ \\
33  &  NGC4565 & $ 54.0\pm  0.1$ & $ 56.6\pm  0.2$ & $ 94.8\pm  0.3$ & $ 92.2\pm  1.3$ & $ 199.6\pm   0.1$ & $ 200.6\pm   0.3$ & $ 222.3\pm   0.8$ & $ 222.3\pm   3.3$ \\
34  &  NGC4569 & $ 38.8\pm  0.1$ & $ 51.0\pm  0.2$ & $ 28.8\pm  0.3$ & $ 23.2\pm  0.7$ & $ 151.9\pm   0.2$ & $ 162.9\pm   0.8$ & $  94.6\pm   1.9$ & $  85.6\pm   4.0$ \\
35  &  NGC4594 & $ 35.3\pm  0.1$ & $ 36.6\pm  0.1$ & $ 68.3\pm  0.9$ & $ 71.6\pm  5.1$ & $ 143.7\pm   0.1$ & $ 145.4\pm   0.2$ & $ 163.0\pm   1.8$ & $ 189.2\pm  10.1$ \\
36  &  NGC4631 & $ 54.4\pm  0.2$ & $ 52.3\pm  0.2$ & $ 52.7\pm  0.2$ & $ 45.5\pm  0.2$ & $ 196.9\pm   0.2$ & $ 194.2\pm   0.5$ & $ 191.5\pm   0.5$ & $ 188.8\pm   1.1$ \\
37  &  NGC4666 & $ 18.4\pm  0.1$ & $ 18.5\pm  0.1$ & $ 24.3\pm  0.1$ & $ 25.4\pm  0.1$ & $  71.1\pm   0.1$ & $  71.0\pm   0.1$ & $  77.5\pm   0.1$ & $  74.1\pm   0.3$ \\
38  &  NGC4710 & $ 19.9\pm  0.1$ & $ 17.8\pm  0.1$ & $  9.5\pm  0.1$ & $ 10.2\pm  0.2$ & $  76.3\pm   0.1$ & $  70.0\pm   0.3$ & $  50.5\pm   3.2$ & $  40.5\pm   3.7$ \\
39  &  NGC5102 & $ 25.1\pm  0.1$ & $ 24.5\pm  0.1$ & $ 24.4\pm  0.8$ & $ 21.1\pm  2.9$ & $ 114.1\pm   0.1$ & $ 113.0\pm   0.1$ & $ 113.9\pm   2.2$ & $ 125.4\pm   5.8$ \\
40  &  NGC5170 & $ 31.7\pm  0.1$ & $ 33.3\pm  0.2$ & $ 81.1\pm  0.8$ & $ 90.6\pm  2.7$ & $ 143.7\pm   0.2$ & $ 146.1\pm   0.4$ & $ 157.5\pm   1.2$ & $ 163.9\pm   4.8$ \\
41  &  NGC5253 & $ 15.9\pm  0.1$ & $  6.0\pm  0.1$ & $  6.8\pm  0.1$ & $  9.0\pm  0.1$ & $  78.2\pm   0.1$ & $  56.9\pm   0.2$ & $  28.6\pm   0.1$ & $  29.9\pm   0.1$ \\
42  &  NGC5422 & $  9.1\pm  0.1$ & $ 10.3\pm  0.1$ & $ 11.6\pm  0.6$ & $ 17.8\pm  3.8$ & $  38.7\pm   0.2$ & $  47.3\pm   0.2$ & $  50.9\pm   2.3$ & $  60.6\pm   5.1$ \\
43  &  NGC5746 & $ 28.6\pm  0.1$ & $ 29.8\pm  0.1$ & $ 59.9\pm  0.3$ & $ 57.9\pm  1.2$ & $ 128.1\pm   0.1$ & $ 127.3\pm   0.2$ & $ 134.1\pm   0.7$ & $ 132.2\pm   3.5$ \\
44  &  NGC5775 & $ 22.2\pm  0.1$ & $ 23.1\pm  0.1$ & $ 31.7\pm  0.1$ & $ 32.8\pm  0.1$ & $  74.4\pm   0.1$ & $  74.8\pm   0.1$ & $  77.3\pm   0.1$ & $  80.3\pm   0.1$ \\
45  &  NGC5866 & $ 17.5\pm  0.1$ & $ 17.7\pm  0.1$ & $ 16.5\pm  0.1$ & $ 18.9\pm  0.6$ & $  80.9\pm   0.1$ & $  80.1\pm   0.1$ & $  66.4\pm   1.2$ & $  83.9\pm   7.4$ \\
46  &  NGC6503 & $ 27.4\pm  0.1$ & $ 27.6\pm  0.1$ & $ 31.1\pm  0.1$ & $ 33.1\pm  0.1$ & $  85.2\pm   0.1$ & $  85.1\pm   0.1$ & $  76.5\pm   0.1$ & $  72.8\pm   0.4$ \\
47  &  NGC6764 & $ 15.1\pm  0.1$ & $ 14.9\pm  0.1$ & $  8.9\pm  0.1$ & $  9.2\pm  0.1$ & $  50.5\pm   0.1$ & $  50.2\pm   0.1$ & $  46.4\pm   0.1$ & $  34.7\pm   0.3$ \\
48  &  NGC7090 & $ 53.3\pm  0.2$ & $ 49.8\pm  2.5$ & $ 34.1\pm  0.4$ & $ 29.9\pm  1.0$ & $ 138.7\pm   0.3$ & $ 126.2\pm   0.9$ & $ 111.5\pm   1.6$ & $ 106.5\pm   5.7$ \\
49  &  NGC7457 & $ 20.2\pm  0.1$ & $ 20.8\pm  0.1$ & $ 28.9\pm  1.5$ & $ 48.2\pm  6.0$ & $  73.8\pm   0.1$ & $  71.4\pm   0.4$ & $  92.2\pm   4.4$ & $ 121.9\pm   6.8$ \\
50  &  NGC7582 & $ 10.7\pm  0.1$ & $  5.9\pm  0.1$ & $  9.3\pm  0.1$ & $ 10.5\pm  0.1$ & $  70.4\pm   0.1$ & $  58.4\pm   0.1$ & $  61.8\pm   0.1$ & $  48.7\pm   0.2$ \\
51  &  NGC7814 & $ 14.8\pm  0.1$ & $ 15.6\pm  0.1$ & $ 21.2\pm  0.8$ & $ 38.1\pm  6.2$ & $  91.2\pm   0.2$ & $  88.5\pm   0.5$ & $ 110.4\pm   4.1$ & $ 151.1\pm   8.3$ \\
\enddata
\tablecomments{\scriptsize 1--8: The radii enclosed 50\% and 90\% of the total luminosity for \wise~each band. 9: The inclination angle obtained from the disk fitting (1 $\sigma$ error is 0.01)}%9--12: The integrated fluxes from photometric measurements.  Due to the error pixels in the \wise image data, there is no record of M82.}%The aperture for bands W1, W2, W3, and W4 is based on the radius observed under 2.17 microns at total isophotal level published in NED.
\label{tab:con}
\end{deluxetable*}

\begin{deluxetable*}{lcccccccc}
\centering
\tiny
  \tabletypesize{\tiny}
  \tablecaption{Result of the sample galaxies}
  \tablewidth{0pt}
  \tablehead{
 \colhead{Name} & \colhead{$Flux (3.4\mu m)$} & \colhead{$Flux (4.6\mu m)$} & \colhead{$Flux (12\mu m)$} & \colhead{$Flux (22\mu m)$} & \colhead{$M_\star^1$} & \colhead{$M_\star^2$} & \colhead{$\sfr_{IR}$ $(12\mu m)$} & \colhead{$\sfr_{IR}$ $(22\mu m)$} \\
 & (Jy) & (Jy) & (Jy) & (Jy) & ($10^{10}M_\odot$) & ($10^{10}M_\odot$) & $M_\odot yr^{-1}$ & $M_\odot yr^{-1}$ \\
 & (1) & (2) & (3) & (4) & (5) & (6) & (7) & (8)
}
\startdata
IC2560   & $ 0.123 \pm 0.002  $ & $  0.085 \pm 0.001   $ & $   0.31 \pm  0.01  $ & $    1.04  \pm  0.02   $ & $     1.19 \pm   0.05   $ &  $     2.12 \pm   0.01  $ & $  0.93  \pm  0.02 $ & $  2.61  \pm  0.05  $ \\
NGC24    & $ 0.104 \pm 0.002  $ & $  0.059 \pm 0.001   $ & $  0.094 \pm 0.002  $ & $    0.12  \pm  0.01   $ & $     0.26 \pm   0.01   $ &  $    0.212 \pm  0.001  $ & $ 0.028  \pm 0.001 $ & $ 0.030  \pm 0.002  $ \\
NGC520   & $ 0.180 \pm 0.003  $ & $  0.131 \pm 0.002   $ & $   0.71 \pm  0.01  $ & $    2.34  \pm  0.04   $ & $     1.23 \pm   0.06   $ &  $     2.60 \pm   0.01  $ & $  1.93  \pm  0.04 $ & $  5.31  \pm  0.09  $ \\
NGC660   & $  0.55 \pm  0.01  $ & $   0.52 \pm  0.01   $ & $   2.30 \pm  0.04  $ & $    6.76  \pm  0.12   $ & $     0.26 \pm   0.01   $ &  $     2.32 \pm   0.01  $ & $  1.75  \pm  0.03 $ & $  4.30  \pm  0.07  $ \\
NGC891   & $  1.96 \pm  0.03  $ & $   1.34 \pm  0.02   $ & $   4.81 \pm  0.09  $ & $    6.86  \pm  0.12   $ & $     2.33 \pm   0.10   $ &  $     3.96 \pm   0.01  $ & $  1.68  \pm  0.03 $ & $  2.00  \pm  0.04  $ \\
NGC1023  & $  1.28 \pm  0.02  $ & $   0.71 \pm  0.01   $ & $   0.20 \pm  0.01  $ & $    0.07  \pm  0.03   $ & $     6.38 \pm   0.29   $ &  $     6.24 \pm   0.04  $ & $ 0.105  \pm 0.003 $ & $  0.03  \pm  0.01  $ \\
NGC1380  & $  0.63 \pm  0.01  $ & $   0.35 \pm  0.01   $ & $  0.121 \pm 0.003  $ & $    0.10  \pm  0.01   $ & $   10.16  \pm  0.45   $ &  $     9.84  \pm  0.04   $ & $ 0.209  \pm 0.005 $ & $  0.14  \pm  0.01  $ \\
NGC1386  & $ 0.220 \pm 0.003  $ & $  0.180 \pm 0.003   $ & $   0.48 \pm  0.01  $ & $    1.46  \pm  0.03   $ & $     0.25 \pm   0.01   $ &  $    1.133 \pm  0.004  $ & $  0.40  \pm  0.01 $ & $  1.00  \pm  0.02  $ \\
NGC1482  & $ 0.200 \pm 0.003  $ & $  0.152 \pm 0.002   $ & $   1.18 \pm  0.02  $ & $    3.69  \pm  0.06   $ & $     0.53 \pm   0.02   $ &  $    1.317 \pm  0.005  $ & $  1.60  \pm  0.03 $ & $  4.16  \pm  0.07  $ \\
NGC1808  & $  0.92 \pm  0.01  $ & $   0.63 \pm  0.01   $ & $   4.14 \pm  0.08  $ & $  16.08   \pm 0.28   $ & $     1.66  \pm  0.07   $ &  $     2.48  \pm  0.01   $ & $  2.21  \pm  0.04 $ & $  7.15  \pm  0.12  $ \\
NGC2787  & $  0.44 \pm  0.01  $ & $  0.241 \pm 0.004   $ & $  0.080 \pm 0.002  $ & $    0.04  \pm  0.01   $ & $     2.94 \pm   0.13   $ &  $     2.60 \pm   0.01  $ & $ 0.052  \pm 0.001 $ & $ 0.024  \pm 0.004  $ \\
NGC2841  & $  1.35 \pm  0.02  $ & $   0.76 \pm  0.01   $ & $   0.97 \pm  0.02  $ & $    1.02  \pm  0.03   $ & $     9.46 \pm   0.43   $ &  $     7.01 \pm   0.03  $ & $  0.68  \pm  0.01 $ & $  0.59  \pm  0.02  $ \\
NGC3079  & $  0.56 \pm  0.01  $ & $   0.39 \pm  0.01   $ & $   1.85 \pm  0.04  $ & $    3.08  \pm  0.05   $ & $     1.72 \pm   0.08   $ &  $     2.91 \pm   0.01  $ & $  1.78  \pm  0.03 $ & $  2.47  \pm  0.04  $ \\
NGC3115  & $  1.44 \pm  0.02  $ & $   0.79 \pm  0.01   $ & $   0.21 \pm  0.01  $ & $    0.28  \pm  0.04   $ & $     5.19 \pm   0.23   $ &  $     5.09 \pm   0.04  $ & $ 0.078  \pm 0.003 $ & $  0.08  \pm  0.01  $ \\
NGC3198  & $ 0.281 \pm 0.004  $ & $  0.172 \pm 0.003   $ & $   0.61 \pm  0.01  $ & $    1.13  \pm  0.03   $ & $     1.29 \pm   0.06   $ &  $    1.217 \pm  0.005  $ & $  0.45  \pm  0.01 $ & $  0.70  \pm  0.02  $ \\
NGC3384  & $  0.66 \pm  0.01  $ & $   0.36 \pm  0.01   $ & $  0.087 \pm 0.004  $ & $    0.07  \pm  0.02   $ & $     3.59 \pm   0.16   $ &  $     3.49 \pm   0.03  $ & $ 0.047  \pm 0.002 $ & $  0.03  \pm  0.01  $ \\
NGC3412  & $ 0.299 \pm 0.005  $ & $  0.164 \pm 0.003   $ & $  0.043 \pm 0.002  $ & $    0.03  \pm  0.01   $ & $     1.56 \pm   0.07   $ &  $     1.47 \pm   0.01  $ & $ 0.022  \pm 0.001 $ & $ 0.011  \pm 0.004  $ \\
NGC3521  & $  1.94 \pm  0.03  $ & $   1.20 \pm  0.02   $ & $   4.91 \pm  0.09  $ & $    6.36  \pm  0.11   $ & $     5.08 \pm   0.23   $ &  $     4.83 \pm   0.02  $ & $  2.17  \pm  0.04 $ & $  2.34  \pm  0.04  $ \\
NGC3556  & $  0.66 \pm  0.01  $ & $   0.43 \pm  0.01   $ & $   2.11 \pm  0.04  $ & $    4.19  \pm  0.07   $ & $     1.09 \pm   0.05   $ &  $     1.43 \pm   0.01  $ & $  0.85  \pm  0.02 $ & $  1.41  \pm  0.03  $ \\
NGC3628  & $  1.47 \pm  0.02  $ & $   0.92 \pm  0.01   $ & $   2.84 \pm  0.05  $ & $    2.61  \pm  0.06   $ & $     4.76 \pm   0.21   $ &  $     5.36 \pm   0.02  $ & $  1.72  \pm  0.03 $ & $  1.32  \pm  0.03  $ \\
NGC3877  & $ 0.307 \pm 0.005  $ & $  0.189 \pm 0.003   $ & $   0.67 \pm  0.01  $ & $    1.04  \pm  0.02   $ & $     1.29 \pm   0.06   $ &  $    1.257 \pm  0.004  $ & $  0.47  \pm  0.01 $ & $  0.61  \pm  0.01  $ \\
NGC3955  & $ 0.126 \pm 0.002  $ & $  0.080 \pm 0.001   $ & $   0.38 \pm  0.01  $ & $    0.70  \pm  0.01   $ & $     0.92 \pm   0.04   $ &  $    1.027 \pm  0.004  $ & $  0.56  \pm  0.01 $ & $  0.87  \pm  0.02  $ \\
NGC3957  & $ 0.106 \pm 0.002  $ & $  0.060 \pm 0.001   $ & $  0.046 \pm 0.001  $ & $   0.051  \pm 0.003   $ & $     2.65 \pm   0.12   $ &  $     2.35 \pm   0.01  $ & $ 0.123  \pm 0.003 $ & $  0.11  \pm  0.01  $ \\
NGC4013  & $  0.36 \pm  0.01  $ & $  0.225 \pm 0.004   $ & $   0.52 \pm  0.01  $ & $    0.69  \pm  0.01   $ & $     2.44 \pm   0.11   $ &  $     2.92 \pm   0.01  $ & $  0.65  \pm  0.01 $ & $  0.73  \pm  0.01  $ \\
NGC4111  & $  0.33 \pm  0.01  $ & $  0.183 \pm 0.003   $ & $  0.072 \pm 0.002  $ & $    0.09  \pm  0.01   $ & $     2.71 \pm   0.12   $ &  $     2.53 \pm   0.01  $ & $ 0.062  \pm 0.002 $ & $ 0.068  \pm 0.004  $ \\
NGC4217  & $  0.43 \pm  0.01  $ & $  0.283 \pm 0.004   $ & $   1.02 \pm  0.02  $ & $    1.37  \pm  0.03   $ & $     2.47 \pm   0.11   $ &  $     3.36 \pm   0.01  $ & $  1.37  \pm  0.03 $ & $  1.53  \pm  0.03  $ \\
NGC4244  & $ 0.310 \pm 0.005  $ & $  0.183 \pm 0.003   $ & $   0.21 \pm  0.01  $ & $    0.34  \pm  0.03   $ & $     0.15 \pm   0.01   $ &  $    0.158 \pm  0.001  $ & $ 0.014  \pm 0.001 $ & $ 0.019  \pm 0.001  $ \\
NGC4251  & $ 0.268 \pm 0.004  $ & $  0.149 \pm 0.002   $ & $  0.043 \pm 0.001  $ & $    0.03  \pm  0.01   $ & $     3.72 \pm   0.17   $ &  $     3.75 \pm   0.02  $ & $ 0.064  \pm 0.002 $ & $  0.04  \pm  0.01  $ \\
NGC4388  & $ 0.280 \pm 0.004  $ & $  0.232 \pm 0.004   $ & $   0.72 \pm  0.01  $ & $    2.70  \pm  0.05   $ & $     0.36 \pm   0.02   $ &  $     1.74 \pm   0.01  $ & $  0.74  \pm  0.01 $ & $  2.32  \pm  0.04  $ \\
NGC4438  & $  0.48 \pm  0.01  $ & $  0.271 \pm 0.004   $ & $   0.26 \pm  0.01  $ & $    0.27  \pm  0.01   $ & $     3.12 \pm   0.14   $ &  $     2.75 \pm   0.01  $ & $ 0.189  \pm 0.004 $ & $  0.16  \pm  0.01  $ \\
NGC4501  & $  1.16 \pm  0.02  $ & $   0.69 \pm  0.01   $ & $   1.91 \pm  0.04  $ & $    2.55  \pm  0.05   $ & $     7.31 \pm   0.33   $ &  $     6.21 \pm   0.02  $ & $  1.66  \pm  0.03 $ & $  1.85  \pm  0.04  $ \\
NGC4526  & $  0.93 \pm  0.01  $ & $   0.52 \pm  0.01   $ & $   0.28 \pm  0.01  $ & $    0.40  \pm  0.03   $ & $     9.40 \pm   0.43   $ &  $     8.69 \pm   0.04  $ & $  0.32  \pm  0.01 $ & $  0.37  \pm  0.02  $ \\
NGC4565  & $  1.56 \pm  0.02  $ & $   0.92 \pm  0.01   $ & $   1.58 \pm  0.03  $ & $    1.87  \pm  0.04   $ & $     5.12 \pm   0.23   $ &  $     4.67 \pm   0.02  $ & $  0.69  \pm  0.01 $ & $  0.68  \pm  0.01  $ \\
NGC4569  & $  0.82 \pm  0.01  $ & $   0.51 \pm  0.01   $ & $   0.99 \pm  0.02  $ & $    1.48  \pm  0.04   $ & $     1.58 \pm   0.07   $ &  $     1.89 \pm   0.01  $ & $  0.34  \pm  0.01 $ & $  0.42  \pm  0.01  $ \\
NGC4594  & $  3.81 \pm  0.06  $ & $   2.13 \pm  0.03   $ & $   1.04 \pm  0.02  $ & $    0.89  \pm  0.05   $ & $    12.65 \pm   0.57   $ &  $   11.71  \pm  0.05   $ & $  0.38  \pm  0.01 $ & $  0.27  \pm  0.02  $ \\
NGC4631  & $  1.22 \pm  0.02  $ & $   0.84 \pm  0.01   $ & $   4.68 \pm  0.09  $ & $    7.89  \pm  0.14   $ & $     0.84 \pm   0.04   $ &  $    1.313 \pm  0.005  $ & $  0.96  \pm  0.02 $ & $  1.35  \pm  0.02  $ \\
NGC4666  & $  0.65 \pm  0.01  $ & $   0.44 \pm  0.01   $ & $   2.45 \pm  0.05  $ & $    3.59  \pm  0.06   $ & $     2.11 \pm   0.09   $ &  $     2.97 \pm   0.01  $ & $  2.13  \pm  0.04 $ & $  2.60  \pm  0.05  $ \\
NGC4710  & $ 0.323 \pm 0.005  $ & $  0.180 \pm 0.003   $ & $  0.208 \pm 0.004  $ & $    0.46  \pm  0.01   $ & $     3.17 \pm   0.14   $ &  $     2.43 \pm   0.01  $ & $ 0.207  \pm 0.004 $ & $  0.38  \pm  0.01  $ \\
NGC5102$^a$  & $  0.50 \pm  0.01  $ & $  0.285 \pm 0.004   $ & $  0.088 \pm 0.002  $ & $    0.09  \pm  0.01   $ & $     0.17 \pm   0.01   $ &  $    0.183 \pm  0.001  $ & $ 0.003  \pm 0.001 $ & $ 0.003  \pm 0.001  $ \\
NGC5170  & $  0.34 \pm  0.01  $ & $  0.196 \pm 0.003   $ & $   0.25 \pm  0.01  $ & $    0.27  \pm  0.01   $ & $     4.92 \pm   0.22   $ &  $     4.46 \pm   0.02  $ & $  0.44  \pm  0.01 $ & $  0.40  \pm  0.02  $ \\
NGC5253  & $ 0.252 \pm 0.004  $ & $  0.292 \pm 0.005   $ & $   2.11 \pm  0.04  $ & $  12.82   \pm 0.22   $ & $    0.003  \pm 0.001   $ &  $    0.072  \pm 0.001   $ & $ 0.123  \pm 0.002 $ & $  0.62  \pm  0.01  $ \\
NGC5422  & $ 0.102 \pm 0.002  $ & $  0.056 \pm 0.001   $ & $  0.016 \pm 0.001  $ & $   0.011  \pm 0.003   $ & $     3.82 \pm   0.17   $ &  $     3.60 \pm   0.03  $ & $ 0.058  \pm 0.002 $ & $  0.04  \pm  0.01  $ \\
NGC5746  & $  0.65 \pm  0.01  $ & $   0.37 \pm  0.01   $ & $   0.44 \pm  0.01  $ & $    0.47  \pm  0.01   $ & $   12.48  \pm  0.56   $ &  $   10.59   \pm 0.04    $ & $  0.94  \pm  0.02 $ & $  0.84  \pm  0.02  $ \\
NGC5775  & $  0.34 \pm  0.01  $ & $  0.233 \pm 0.004   $ & $   1.22 \pm  0.02  $ & $    1.88  \pm  0.03   $ & $     2.94 \pm   0.13   $ &  $     4.55 \pm   0.02  $ & $  3.06  \pm  0.06 $ & $  3.93  \pm  0.07  $ \\
NGC5866  & $  0.69 \pm  0.01  $ & $   0.39 \pm  0.01   $ & $   0.26 \pm  0.01  $ & $    0.26  \pm  0.01   $ & $     4.83 \pm   0.22   $ &  $     4.89 \pm   0.02  $ & $ 0.215  \pm 0.004 $ & $  0.18  \pm  0.01  $ \\
NGC6503  & $  0.46 \pm  0.01  $ & $  0.278 \pm 0.004   $ & $   0.78 \pm  0.01  $ & $    0.97  \pm  0.02   $ & $     0.29 \pm   0.01   $ &  $    0.277 \pm  0.001  $ & $ 0.076  \pm 0.002 $ & $ 0.079  \pm 0.001  $ \\
NGC6764  & $ 0.083 \pm 0.001  $ & $  0.054 \pm 0.001   $ & $   0.32 \pm  0.01  $ & $    1.22  \pm  0.02   $ & $     0.90 \pm   0.04   $ &  $    1.038 \pm  0.004  $ & $  0.78  \pm  0.01 $ & $  2.46  \pm  0.04  $ \\
NGC7090  & $ 0.235 \pm 0.004  $ & $  0.140 \pm 0.002   $ & $   0.42 \pm  0.01  $ & $    0.71  \pm  0.02   $ & $     0.22 \pm   0.01   $ &  $    0.198 \pm  0.001  $ & $ 0.059  \pm 0.001 $ & $ 0.083  \pm 0.003  $ \\
NGC7457  & $ 0.201 \pm 0.003  $ & $  0.113 \pm 0.002   $ & $  0.044 \pm 0.002  $ & $    0.04  \pm  0.01   $ & $     1.19 \pm   0.06   $ &  $     1.19 \pm   0.01  $ & $ 0.029  \pm 0.001 $ & $  0.02  \pm  0.01  $ \\
NGC7582  & $  0.56 \pm  0.01  $ & $   0.54 \pm  0.01   $ & $   1.94 \pm  0.04  $ & $    6.89  \pm  0.12   $ & $     0.59 \pm   0.03   $ &  $     6.11 \pm   0.02  $ & $  3.62  \pm  0.07 $ & $10.72  \pm  0.19  $ \\
NGC7814  & $  0.52 \pm  0.01  $ & $  0.290 \pm 0.005   $ & $  0.181 \pm 0.005  $ & $    0.13  \pm  0.02   $ & $     5.84 \pm   0.27   $ &  $     5.22 \pm   0.03  $ & $  0.21  \pm  0.01 $ & $  0.13  \pm  0.02  $ \\

\enddata
	\tablecomments{\scriptsize (1$\sim$ 4) The flux measured in \wise~within the galaxies aperture. (5$\sim$6) The estimated SMs derived from Equation~\ref{equi:W1sm} and Equation~\ref{equi:W2sm}.  (7$\sim$ 8) The estimated \sfrs based on the $12\mu m$ and $22\mu m$, respectively.  a: The photometry aperture is reduced by half in the W3 and W4 of NGC5102.}
\label{tab:sfrsm}
\end{deluxetable*}

\begin{deluxetable*}{lcccccccc}
\centering
\tiny
  \tabletypesize{\tiny}
  \tablecaption{The parameters of X-ray corona}
  \tablewidth{0pt}
  \tablehead{
	  %\colhead{Name} & \colhead{$L_X$} & \colhead{$H_{50}$} & \colhead{$H_{75}$} & \colhead{$H_{95}$} & \colhead{$H_5$} \\
	  \colhead{Name} & \colhead{$L_X$} & \colhead{$H_{50+}$} & \colhead{$H_{50-}$} & \colhead{$H_{75+}$} & \colhead{$H_{75-}$} & \colhead{$H_{95+}$} & \colhead{$H_{95-}$} & \colhead{i} \\
	& $(10^{38} erg~s^{-1})$& (kpc) & (kpc) & (kpc) & (kpc) & (kpc) & (kpc) & (degree) \\
	& (1) & (2) & (3) & (4) & (5) & (6) & (7) & (8)
}
\startdata
IC2560  & 	 $115.2\pm7.0$ 	         & $ 2.89\pm 0.51$&$ 4.67\pm 0.34$&$13.76\pm 1.02$&$ 9.43\pm 0.59$&$20.90\pm 0.93$&$19.54\pm 1.10$  &   62.89 \\
NGC0024 & 	 $1.7\pm0.3$ 	         & $ 1.80\pm 1.45$&$ 6.07\pm 1.82$&$ 3.25\pm 1.14$&$10.01\pm 1.66$&$ 3.86\pm 0.45$&      ---        &   $74.91\pm0.03$ \\
NGC0520 & 	 $19.4_{-6.3}^{+3.8}$ 	 & $ 0.89\pm 0.16$&$ 1.29\pm 0.16$&$ 2.43\pm 0.73$&$ 2.99\pm 0.40$&$14.64\pm 3.23$&$12.86\pm 6.55$  &   59.01 \\
NGC0660 & 	 $11.8_{-2.0}^{+3.1}$ 	 & $ 0.21\pm 0.09$&$ 1.11\pm 0.38$&$ 0.68\pm 0.38$&$ 2.74\pm 0.60$&$ 4.32\pm 3.63$&$11.42\pm 3.59$  &   69.51 \\
NGC0891 & 	 $38.0\pm0.9$ 	         & $ 2.20\pm 0.03$&$ 1.91\pm 0.03$&$ 4.98\pm 0.12$&$ 5.24\pm 0.17$&$ 9.96\pm 0.12$&$10.54\pm 0.12$  &   85.75 \\
NGC1023 & 	 $2.9_{-0.7}^{+0.6}$ 	 & $ 1.11\pm 0.07$&$ 0.74\pm 0.03$&$ 3.31\pm 0.98$&$ 1.99\pm 0.78$&      ---      &      ---        &   68.78 \\
NGC1380 & 	 $40.1_{-3.8}^{+3.5}$ 	 & $ 0.86\pm 0.06$&$ 0.68\pm 0.06$&$ 3.08\pm 0.49$&$ 1.48\pm 0.12$&$15.85\pm 3.70$&$10.24\pm 5.06$  &   57.73 \\
NGC1386 & 	 $15.0_{-2.0}^{+1.2}$ 	 & $ 1.25\pm 0.58$&$ 0.71\pm 0.13$&$ 5.56\pm 2.00$&$ 3.69\pm 0.67$&$11.44\pm 1.20$&$ 9.70\pm 0.49$  &   64.47 \\
NGC1482 & 	 $75.1_{-6.7}^{+6.2}$ 	 & $ 0.51\pm 0.06$&$ 1.25\pm 0.06$&$ 1.25\pm 0.17$&$ 2.45\pm 0.17$&$13.23\pm 5.25$&$ 8.10\pm 1.37$  &   51.75 \\
NGC1808 & 	 $25.8_{-2.9}^{+2.6}$ 	 & $ 0.29\pm 0.00$&$ 0.25\pm 0.04$&$ 0.50\pm 0.04$&$ 1.00\pm 0.07$&$ 1.79\pm 0.11$&$ 9.34\pm 2.47$  &   66.18 \\
NGC2787 & 	 $1.8_{-1.3}^{+1.9}$ 	 & $11.50\pm 0.23$&$ 0.61\pm 0.23$&$11.50\pm 0.19$&$ 1.55\pm 0.49$&$11.42\pm 0.19$&$ 3.44\pm 0.53$  &   49.38 \\
NGC2841 & 	 $19.8_{-3.2}^{+2.8}$ 	 & $ 1.64\pm 0.12$&$ 1.39\pm 0.21$&$ 3.77\pm 0.25$&$ 5.29\pm 0.45$&$12.10\pm 0.98$&$12.88\pm 0.53$  &   64.14 \\
NGC3079 & 	 $85.9_{-4.9}^{+4.7}$ 	 & $ 1.20\pm 0.05$&$ 3.22\pm 0.14$&$ 3.70\pm 0.14$&$ 6.96\pm 0.24$&$11.04\pm 0.43$&$14.25\pm 1.01$  &   77.67 \\
NGC3115 & 	 $0.4_{-0.2}^{+0.1}$ 	 & $ 0.82\pm 0.09$&$ 0.63\pm 0.06$&$10.12\pm 0.28$&$ 9.15\pm 0.71$&      ---      &      ---        &   76.14 \\
NGC3198 & 	 $16.7\pm1.8$ 	         & $ 5.65\pm 1.81$&$ 4.34\pm 2.70$&$ 3.75\pm 0.72$&$ 3.67\pm 0.46$&      ---      &      ---        &   68.89 \\
NGC3384 & 	 $12.6\pm2.2$ 	         & $ 0.82\pm 1.75$&$ 1.13\pm 2.75$&$ 3.23\pm 1.13$&$ 4.05\pm 1.00$&$ 5.77\pm 0.55$&$ 5.53\pm 0.89$  &   58.41 \\
NGC3412 & 	 $9.8\pm1.1$ 	         & $ 8.13\pm 3.18$&$ 7.79\pm 1.57$&$ 7.83\pm 3.45$&$ 7.79\pm 1.94$&$ 7.53\pm 3.58$&$ 8.70\pm 1.27$  &   55.18 \\
NGC3521 & 	 $26.3\pm1.8$ 	         & $ 1.76\pm 0.10$&$ 1.30\pm 0.07$&$ 3.29\pm 0.16$&$ 2.61\pm 0.16$&$ 9.58\pm 1.53$&$ 9.35\pm 2.25$  &   63.45 \\
NGC3556 & 	 $13.1_{-2.6}^{+2.3}$ 	 & $ 0.87\pm 0.03$&$ 1.21\pm 0.03$&$ 2.27\pm 0.16$&$ 2.24\pm 0.12$&      ---      &$11.39\pm 1.21$  &   77.69 \\
NGC3628 & 	 $38.7_{-3.3}^{+2.9}$ 	 & $ 2.32\pm 0.08$&$ 4.73\pm 0.08$&$ 4.76\pm 0.11$&$ 7.93\pm 0.23$&$ 8.80\pm 0.46$&$12.50\pm 0.38$  &   81.46 \\
NGC3877 & 	 $4.3_{-2.4}^{+2.2}$ 	 & $ 0.98\pm 0.08$&$ 1.15\pm 0.08$&$ 5.04\pm 0.86$&$10.25\pm 2.13$&      ---      &      ---        &   76.52 \\
NGC3955 & 	 $10.9_{-6.3}^{+3.7}$ 	 & $ 0.54\pm 0.30$&$ 0.90\pm 0.18$&$ 1.44\pm 6.05$&$ 1.44\pm 0.30$&$ 6.53\pm 4.07$&$ 6.23\pm 2.46$  &   69.45 \\
NGC3957 & 	 $19.8\pm3.0$ 	         & $ 0.88\pm 0.40$&$ 9.92\pm 4.00$&$ 3.28\pm 2.00$&$18.24\pm 2.48$&$ 7.28\pm 3.76$&$20.40\pm 1.12$  &   73.29 \\
NGC4013 & 	 $23.3\pm2.2$ 	         & $ 1.43\pm 0.16$&$ 4.12\pm 1.59$&$10.39\pm 4.23$&$21.06\pm 1.87$&               &      ---        &   81.27 \\
NGC4111 & 	 $6.7_{-3.8}^{+2.4}$ 	 & $ 2.18\pm 0.22$&$ 3.62\pm 0.35$&$ 8.90\pm 0.48$&$ 9.60\pm 0.65$&$15.23\pm 0.61$&$15.88\pm 0.92$  &   76.35 \\
NGC4217 & 	 $75.9\pm5.9$ 	         & $ 0.85\pm 0.11$&$ 1.76\pm 0.17$&$ 3.01\pm 1.42$&$12.42\pm 3.29$&$16.22\pm 3.29$&      ---        &   82.05 \\
NGC4244 & 	 $1.1\pm0.1$ 	         & $ 0.95\pm 0.85$&$ 2.05\pm 0.57$&$ 2.72\pm 0.89$&$ 4.75\pm 0.57$&      ---      &      ---        &   81.82 \\
NGC4251 & 	 $23.2\pm4.6$ 	         & $12.54\pm 1.60$&$15.22\pm 1.08$&$12.94\pm 1.03$&      ---      &               &      ---        &   62.84 \\
NGC4388 & 	 $73.1_{-9.2}^{+11.5}$ 	 & $ 0.65\pm 0.05$&$ 1.79\pm 0.05$&$ 1.74\pm 0.15$&$ 4.33\pm 0.20$&$ 6.62\pm 1.29$&$15.47\pm 1.09$  &   71.86 \\
NGC4438 & 	 $82.9_{-6.2}^{+5.7}$ 	 & $ 3.94\pm 0.08$&$ 1.76\pm 0.21$&$ 8.13\pm 0.42$&$ 8.00\pm 1.38$&$15.29\pm 0.25$&$16.00\pm 0.59$  &   61.05 \\
NGC4501 & 	 $138.1_{-26.4}^{+25.3}$ & $ 1.60\pm 0.05$&$ 1.74\pm 0.18$&$ 2.88\pm 0.09$&$ 3.93\pm 0.27$&$16.99\pm 0.64$&$12.97\pm 0.46$  &   61.21 \\
NGC4526 & 	 $18.7_{-4.3}^{+3.9}$ 	 & $ 0.55\pm 0.05$&$ 1.00\pm 0.05$&$ 1.60\pm 0.20$&$ 2.35\pm 0.15$&      ---      &$11.41\pm 1.30$  &   70.51 \\
NGC4565 & 	 $10.9_{-1.0}^{+1.1}$ 	 & $ 2.55\pm 0.26$&$ 3.58\pm 0.19$&$ 8.04\pm 0.45$&$ 6.65\pm 0.32$&$11.98\pm 0.23$&$12.40\pm 0.71$  &   84.95 \\
NGC4569 & 	 $23.4_{-9.1}^{+4.0}$ 	 & $ 0.54\pm 0.03$&$ 2.55\pm 0.14$&$ 3.44\pm 0.40$&$ 5.85\pm 0.26$&$10.90\pm 0.43$&$10.41\pm 0.34$  &   67.98 \\
NGC4594 & 	 $39.2_{-2.4}^{+2.0}$ 	 & $ 2.25\pm 0.06$&$ 2.76\pm 0.06$&$ 6.74\pm 0.11$&$ 6.99\pm 0.09$&$14.24\pm 0.14$&$14.41\pm 0.20$  &   80.76 \\
NGC4631 & 	 $36.4_{-1.4}^{+1.3}$ 	 & $ 1.60\pm 0.02$&$ 1.42\pm 0.02$&$ 3.68\pm 0.04$&$ 2.79\pm 0.04$&$ 7.82\pm 0.09$&$ 6.76\pm 0.11$  &   79.27 \\
NGC4666 & 	 $86.8_{-38.5}^{+14.2}$  & $ 0.64\pm 0.14$&$ 1.46\pm 0.55$&$ 1.96\pm 0.41$&$ 3.47\pm 3.24$&$13.34\pm 1.64$&$ 5.02\pm 2.56$  &   75.66 \\
NGC4710 & 	 $6.1_{-3.5}^{+0.8}$ 	 & $ 0.20\pm 1.86$&$ 0.59\pm 0.10$&$ 3.52\pm 5.38$&$ 5.23\pm 3.27$&$ 5.86\pm 4.01$&$15.05\pm 1.22$  &   76.37 \\
NGC5102 & 	 $0.6\pm0.1$ 	         & $ 0.51\pm 0.50$&$ 0.27\pm 0.06$&$ 2.46\pm 0.39$&$ 1.44\pm 0.38$&      ---      &$ 2.15\pm 0.13$  &   67.35 \\
NGC5170 & 	 $32.7\pm8.3$ 	         & $ 2.49\pm 3.80$&$27.95\pm 6.87$&$ 8.57\pm 6.54$&$27.49\pm 0.92$&      ---      &      ---        &   83.85 \\
NGC5253 & 	 $1.8\pm0.1$ 	         & $ 0.22\pm 0.01$&$ 0.19\pm 0.01$&$ 0.73\pm 0.04$&$ 0.62\pm 0.05$&      ---      &      ---        &   57.43 \\
NGC5422 & 	 $18.5\pm3.4$ 	         & $20.31\pm 7.01$&$12.22\pm 4.76$&$27.95\pm 2.52$&$16.54\pm 2.79$&      ---      &$18.61\pm 2.79$  &   75.97 \\
NGC5746 & 	 $17.2_{-10.0}^{+6.4}$ 	 & $ 6.25\pm 3.88$&$ 3.74\pm 8.77$&$23.57\pm 4.96$&$15.81\pm 6.32$&$30.82\pm 1.22$&$22.70\pm 5.96$  &   82.46 \\
NGC5775 & 	 $101.5_{-11.9}^{+10.2}$ & $ 1.71\pm 0.08$&$ 3.50\pm 0.16$&$ 6.06\pm 0.70$&$15.69\pm 1.63$&$29.90\pm 0.39$&$27.80\pm 1.17$  &   80.13 \\
NGC5866 & 	 $13.7_{-2.4}^{+2.0}$ 	 & $ 0.67\pm 0.04$&$ 0.80\pm 0.04$&$ 1.91\pm 0.18$&$ 1.60\pm 0.18$&$13.04\pm 2.45$&$ 6.90\pm 1.38$  &   65.92 \\
NGC6503 & 	 $1.6\pm0.2$ 	         & $ 0.26\pm 0.26$&$ 0.48\pm 0.15$&$ 1.29\pm 0.87$&$ 2.70\pm 1.23$&$ 2.07\pm 1.20$&$ 5.64\pm 0.44$  &   71.05 \\
NGC6764 & 	 $187.3_{-78.5}^{+53.9}$ & $ 0.30\pm 0.08$&$ 0.46\pm 0.08$&$ 0.84\pm 0.23$&$ 0.76\pm 0.08$&$19.51\pm 4.73$&$15.47\pm 5.41$  &   63.18 \\
NGC7090 & 	 $0.4_{-0.4}^{+0.3}$ 	 & $ 2.25\pm 1.15$&$ 0.90\pm 1.33$&$ 6.59\pm 0.31$&$ 4.90\pm 1.30$&      ---      &      ---        &   79.42 \\
NGC7457 & 	 $5.0\pm1.3$ 	         & $ 7.56\pm 2.84$&$ 7.53\pm 2.30$&$ 9.48\pm 2.19$&$12.71\pm 2.57$&$ 8.95\pm 4.68$&$15.05\pm 1.65$  &   56.66 \\
NGC7582 & 	 $102.3_{-19.3}^{+17.1}$ & $ 0.40\pm 0.07$&$ 0.74\pm 0.07$&$ 1.20\pm 0.13$&$ 1.87\pm 0.33$&$ 9.30\pm 2.88$&$15.39\pm 4.82$  &   71.49 \\
NGC7814 & 	 $28.9\pm4.8$ 	         & $ 0.90\pm 0.32$&$ 2.42\pm 4.16$&$ 1.53\pm 0.74$&$ 6.79\pm 2.00$&$ 3.53\pm 0.95$&$ 7.11\pm 1.05$  &   65.14 \\

\enddata
	\tablecomments{X-ray corona parameters: (1) The total luminosity of corona retrieve from \citealt{Li2013a} (2$\sim$7) show the vertical radiuses enclosed 50\%, 75\% and 95\% 0.5$\sim$1.5 keV luminosity along positive and negative sides (``+" and ``-") of galactic disk. (8) show the inclination angle of sample galaxies}
\label{tab:xray}
\end{deluxetable*}

\bibliographystyle{natbib}
\bibliography{ref}

\end{document}